\documentclass[12pt,english]{article}
\usepackage{graphicx}
\usepackage{epstopdf}
  \usepackage{mathtools}
   \usepackage{hyperref}
\usepackage{url}
\usepackage{breakurl}
\usepackage{amsmath}
\usepackage{babel}
\newcommand{\btt}{{\pmb \beta}}

\begin {document}
\begin{center}
\textbf{ \Large
Equivalence between electromagnetic self-energy and self-mass 
    }\\
\vspace{10pt}
\textbf{\textit{Khokonov M.Kh.$^{\mathit{1}}$   }} \\
\vspace{3pt}
\textit{$^{\mathit{}}$\hspace{-2pt} Kabardino-Balkarian State University, Nalchik, Russian Federation}\\
\textsl{e-mail: khokon6@mail.ru}\\
\textbf{\textit{Andersen J.U.$^{\mathit{2 }}$   }} \\
\vspace{3pt}
\textit{$^{\mathit{}}$\hspace{-2pt} Aarhus University, Aarhus, Denmark}\\
\textsl{e-mail: jua@phys.au.dk}\\
\vspace{3pt}
\end{center}

\begin{abstract}
A cornerstone of physics, Maxwell's theory of electromagnetism, apparently contains a fatal flaw. The standard expressions for the electromagnetic field energy and self-mass of an electron of finite extension do not obey Einstein's famous equation, $E=mc^2$, but instead fulfill this relation with a factor 4/3 on the left-hand side. Many famous physicists have contributed to the debate of this so-called 4/3-problem but without arriving at a complete solution. Here, a comprehensive solution is presented. The problem is caused by an incorrect treatment of rigid-body dynamics. Relativistic effects are important even at low velocities and equivalence between electromagnetic field energy and self-mass of the electron is restored when these effects are included properly. In a description of the translational motion of a rigid body by point-particle dynamics, its mechanical energy and momentum must be defined as a sum of the energies and momenta of its parts for fixed time, not in the laboratory as in the standard expressions but in the rest frame of the body, and for consistency of the description, the energy and momentum of the associated field must be defined in the same way.

\end{abstract}

\section{Introduction}

In classical electrodynamics an accelerated charge gives rise to electromagnetic radiation and also to a field
that reacts back on the charge with a so-called self-force. This force can be divided into components that are
even and odd, respectively, under time reversal  and the rate of work done by these force components changes sign or is invariant, respectively, under time reversal. The former provides an inertial force resisting acceleration and the latter accounts for energy loss to radiation. In addition, this odd component of the self-force includes a term that
induces reversible energy exchange with the near field, the so-called acceleration energy or Schott term  
\cite{Schott} page 253, \cite{Dirac}.  Apart from the presence of this term, the above distinction  is analogous to that between reactive  and resistive impedance in an electronic circuit \cite{Schwinger}.

Here our focus is on the inertial self-force, characterized by an electromagnetic  mass.
According to the theory of special relativity its electromagnetic mass should be given by
\begin{equation}
  m_e =  \frac{U_{el}}{c^2}
\label{Uselfmass}
\end{equation}
where $U_{el}$  is the electromagnetic self-energy in the electron's rest frame, 
\begin{equation}
U_{el} = \frac{1}{2} \, \int \int \ \frac{\rho ({\bf r}') \rho ({\bf r})}
{\mid {\bf r}-{\bf r}'\mid} \ dVdV' \ .
\label{Uself}
\end{equation}
Here  $\rho ({\bf r})$ describes the   charge distribution of the electron  and $c$  is the velocity of light in vacuum.  

As we shall show below, a standard calculation of the total self-force in the rest frame of an electron, based on
Maxwell's equations, leads to
\begin{equation}
{\bf K}_{self} = - \frac{4}{3} m_e {\bf \dot{v}} + \frac{2}{3} \frac{q^2}{c^3}
{\bf \ddot{v}} ,
\label{K_s33}
\end{equation}
where $\bf{v}$  is the velocity, differentiation with respect to time is indicated by a dot, and $m_e$ is given by 
Eq.(\ref{Uselfmass}). 
 When the first, inertial term in Eq.(\ref{K_s33}) is moved to the left hand side of the equation of motion,
$M \dot{\bf{v}} = {\bf K}$, where ${\bf K}$ includes an external force, ${\bf K} = {\bf K}_{ext} + {\bf K}_{self}$, 
4/3 $m_e$  can be interpreted as a
correction to the mechanical mass $M$ and  the unexpected factor 4/3 is referred to as ``the 4/3-problem''.
 In spite of its century-long  history this problem is still discussed in the literature as one that is not fully resolved  (see, for example, Ch.16 in \cite{Jackson}).  Also the form of the second term is unexpected because the power of the emitted radiation is proportional to the square of the acceleration according to the Larmor formula. The reason is the presence of the much debated Schott term mentioned above \cite{Gron, Ulrik2017}.

The seriousness of the 4/3-problem is emphasized by Feynman in his famous Lectures on Physics \cite{Feynman}.  After the discussion of special  relativity and Maxwell's theory of electromagnetism he writes: ``But we want to stop for a moment to show you that this tremendous edifice, which is such a beautiful success in explaining so many phenomena, ultimately falls on its face.  There are difficulties associated with the {\it ideas}  of Maxwell's theory which are not solved by and not directly associated with quantum mechanics.'' Also, it turns out that solution of problems in quantum electrodynamics can often be reduced to the solution of the corresponding classical problem \cite{Bai},\cite{KhokonNitta2004}.

In the standard textbook by Jackson \cite{Jackson} it is argued that this violation of equivalence between mass and energy of
an electron  is a consequence of the fact that the electromagnetic contributions to the energy and momentum do not transform properly (as a four-vector) but that the problem can be removed by  inclusion of non-electromagnetic forces
(Poincar\'e  stresses) required to stabilize the charge \cite{Poincare}.  This  inclusion gives a total divergence-free energy-momentum tensor (named `the stress tensor' in \cite{Jackson})  and hence the correct energy-momentum transformation properties.    Such a mo\-del was proposed by Schwinger  \cite{Schwinger1983}.  

At the end of the last century Rohrlich described the state of the problem under discussion optimistically:
``Returning to the overview of classical charged particle dynamics, one can summarize the present situation as
very satisfactory: for a charged sphere there now exist equations of motion, both relativistic and nonrelativistic, that make sense and that are free of the problems that have plagued the theory for most of this century''  \cite{Rohrlich1997}  (see also the textbook \cite{Rohrlich1990}). 
However, the authors Kalckar, Lindhard and Ulfbeck (KLU) of the paper \cite{Kalckar}, which unfortunately has gone unnoticed by the general physics community and apparently was not known to Rohrlich, did not share this opinion. They stated that ``there is a crucial error in the usual derivations of self-force''  and found  that, after correction of this error, there is complete equivalence between the field energy and the self-mass of an electron and hence no need to introduce Poincar\'e stresses.
 This conclusion was derived from a comprehensive study of the acceleration of a rigid system of charges.  Previously overlooked relativistic corrections associated with Lorentz contraction of a rigid body and time dilation in an accelerated system  turn out to be important even in the limit of velocities much smaller than $c$. 

In the following we shall show how the 4/3-problem can be resolved. 
First we calculate the total self-force in the standard way from the interaction between the elements of charge in   a classical model of the electron.This leads to a  formula  for the inertial self-force and the associated electromagnetic mass with the troublesome factor 4/3.   
However, when relativistic effects are included  \cite{Kalckar}  the 4/3-factor disappears. The key observation is that, owing to Lorentz contraction, different parts of a rigid body must have different accelerations to preserve rigidity and the time intervals required to reach a new velocity are therefore different. This modifies the way forces on different parts of the body should be added. We then demonstrate equivalence from a calculation of the self-force from transport of field momentum across the surface of a sphere surrounding the electron, including a similar relativistic correction. 
 Dirac applied this type of calculation  to a point electron in his  famous 1938-paper 
\cite{Dirac}   but to avoid the problem of infinite self-energy he omitted the reactive term in the self-force.

As an alternative to inclusion of Poincar\'e stresses Rohr\-lich suggested a new definition of the energy-momentum vector of the electromagnetic field around a moving electron, \cite{Rohrlich1990}  Ch. 4, as discussed also in Jackson's textbook, \cite{Jackson} Ch. 16.  Similar solutions were suggested by Fermi already in 1922 \cite{Fermi1922}  and by Wilson \cite{Wilson} and Kwal \cite{Kwal}.
However, the root of the problem with the standard definition remained elusive and introduction of Poincar\'e stresses was considered an alternative option. In Ch. 5  we discuss this (covariant) definition on the basis of the general formalism of the classical theory of fields in Landau and Lifshitz'  textbook, \cite{Landau}.
The papers by Fermi  and Dirac  are discussed in Apps. C and D.

\section{Retarded electromagnetic fields around an electron and the 4/3-problem}

The Maxwell equations for the electric field generated by a moving electron have two solutions, the retarded field,  ${\bf E}_{ret}$, and the advanced field, ${\bf E}_{adv}$, with boundary conditions in the past and in the future, respectively. Normally, only the retarded solution (Li\' enard-Wiechert  field) is considered to have a physical meaning but the advanced field can sometimes be useful because it is connected to the retarded field by time reversal. 

Dirac \cite{Dirac} and Schwinger \cite{Schwinger}  were both only interested in calculating the resistive component of the self-force  which is associated with the component of the retarded field that is odd under time reversal. Schwinger therefore separated the components of the field that are odd and even under time reversal, 
\begin{equation}
{\bf E}_{ret} = \frac{1}{2} ( {\bf E}_{ret} - {\bf E}_{adv} )
+  \frac{1}{2} ( {\bf E}_{ret} + {\bf E}_{adv} )\, ,
\label{Eret1}
\end{equation}
keeping only the first term.  To avoid the problem of infinite self-energy for a point electron,  also Dirac had only retained only the contribution to the self-force from the first term in Eq.(\ref{Eret1}). 
We shall be interested in the total self-force and hence postpone the separation in Eq.(\ref{Eret1}). The rate at
which the electron performs work on the electric field is

\begin{equation}
- \int \ {\bf j} \cdot {\bf E}_{ret} \ dV , 
\label{W}
\end{equation}
where ${\bf j}$   is the current density created by the electron and the integration extends over the whole coordinate space.
This rate is seen to be invariant under time reversal for the first field component in Eq.(\ref{Eret1}) and to change
sign for the second component.

\subsection{Expansion of electromagnetic fields near an accelerated charge}

As shown in App. A, 
the retarded electromagnetic fields in the vicinity of a point charge $q$ are  to second order in the distance $\varepsilon $  from the charge given by
$$
{\bf E}_{ret} \approx \frac{q}{\varepsilon^2 } \, {\bf n} -
\frac{q}{2c\varepsilon } \left[  {\bf n} ({\bf n} \dot{\btt}) +
\dot{\btt} \right] +
$$
\begin{equation}
+ \frac{q}{c^2} \left[ \, \frac{3}{8} \, ({\bf n} \dot{\btt})^2 {\bf n} 
+ \frac{3}{4} \, ({\bf n} \dot{\btt}) \dot{\btt} - \frac{3}{8} \, |\dot{\btt}|^2
{\bf n} +  \frac{2}{3} \, \ddot{\btt} \, \right] \,\,\, , 
\label{Eret_eps}
\end{equation}
\begin{equation}
{\bf H}_{ret} \approx \frac{q}{2c^2 } \, {\bf n} \times \ddot{\btt} \,\,\, .
\label{Hret_eps}
\end{equation}
The velocity of the charge is here assumed to be zero at the time $t$ of observation, $c \beta (t)=0$, and derivatives
with respect to $t$   are indicated by dots. The unit vector ${\bf n}$   points from the position of the charge at time $t$
towards the point of observation.  Note that  to second order in   $\varepsilon$ the magnetic field  does not depend on distance.  The expressions for the advanced fields ${\bf E}_{adv}$ and  ${\bf H}_{adv}$ can be obtained from Eqs. (\ref{Eret_eps}), (\ref{Hret_eps}) by  the substitution $\ddot{\btt} \rightarrow - \ddot{\btt}$.

In Eq. (\ref{Eret_eps}) only the  term proportional to $\ddot{\btt}$ is odd under time reversal and hence the first term in Eq. (\ref{Eret1}) has a finite value at the location of the charge, 
\begin{equation}
\frac{1}{2} \left( {\bf E}_{ret} - {\bf E}_{adv} \right) = \frac{2}{3} \, \frac{q}{c^2} \ddot{\btt} \,\, ,
\label{E_charge}
\end{equation}
and this field multiplied by $q$  gives a damping force,  ${\bf K}_{damp}$,   accounting for irreversible energy loss  but also for reversible energy exchange with the near field (Schott term),
\begin{equation}
{\bf K}_{damp} \cdot \btt c = \frac{2q^2}{3c} \ddot{\btt} \cdot \btt = 
\frac{2q^2}{3c} \left( \frac{d}{dt}  (\dot{\btt}  \btt) - \dot{\btt}^2     \right). 
\label{K_damp}
\end{equation}
Upon integration over time, the first term vanishes for periodic motion or for initial and final states without acceleration while the second term gives a radiation damping in accordance with the Larmor formula for the radiation intensity.
 In contrast to \cite{Dirac}, our aim  is to study not the radiative friction  but  the  electromagnetic self-energy and self-mass  of the electron.  For this purpose only the first two terms in Eq. (\ref{Eret_eps}) are needed.

\subsection{The 4/3-problem for electron field momentum and self-force}

Consider the Abraham-Lorentz classical model of an electron   as a stable spherical shell with radius $R$, on which a total charge $q$  is uniformly distributed \cite{Abraham}, \cite{Lorentz}. In an  inertial frame $K$ the shell moves with velocity $\bf{v}$. According to standard results, the density of momentum of an electromagnetic field equals ${\bf S}/c^2$, where $ {\bf S} = c  {\bf E} \times {\bf H}/(4\pi)$ is the Poynting vector. Introducing  the field belonging to the shell, as observed in the frame $K$, and integrating over all  space one  finds  a momentum \cite{Jackson}, \cite{Feynman},
\begin{equation}
{\bf P}^{(f)} = \frac{1}{c^2} \int dV{\bf S} = \frac{4}{3} \frac{1}{ c^2} \frac{q^2}  {2 R}\gamma {\bf v},
\label{P_el}
\end{equation}
where $\gamma = (1 - \beta^2)^{-1/2}$. 
In the rest frame, $K'$,  the momentum is ${\bf P'}^{(f)} = 0 $  and the energy $E'^{(f)}  = q^2/2R$ (see Eq. (\ref{m_e})), and a Lorentz transformation of this energy-momentum gives Eq.(\ref{P_el}) for the momentum   without the factor 4/3. Thus, due to this offending factor, the energy-momentum fails to transform as a four-vector.

An alternative, more direct way to see the lack of equivalence between electromagnetic self-energy and self-mass is through  a calculation of the electromagnetic self-force of an accelerated charge. 
Following  Heitler   \cite{Heitler},  \S 4, we  consider two charge elements,   $dq$ and $dq'$, on the spherical shell, separated by the distance $\varepsilon$. The charge element $dq$ produces an electric  field acting on the charge $dq'$  with the force  $dq'  d{\bf E}_{ret}$,  where the field is obtained from Eq. (\ref{Eret_eps}) with $q$  replaced by $dq$.
The total force acting  on the electron itself is then equal to
\begin{equation}
{\bf K}_{self} = \int \!\!\! \int \! dq' d{\bf E}_{ret}  .  
\label{K_s}
\end{equation}
Since the unit vector ${\bf n} $ in the expression (\ref{Eret_eps}), pointing from the charge element $dq$  towards $dq'$, is uniformly distributed over solid angles for fixed $\varepsilon$, we obtain the following expression for the self-force \cite{Heitler}:
\begin{equation}
{\bf K}_{self} = \int \!\!\! \int \! dq dq' \int \left[ 
- \frac{{\bf n} ({\bf n} \dot{\btt})}{2c\varepsilon} - 
\frac{ \dot{\btt} }{2c\varepsilon} + \frac{2}{3}
\frac{ \ddot{\btt} }{c^2}
\right] \frac{d \Omega}{4\pi} , 
\label{K_s1}
\end{equation}
where we have omitted odd power terms   with respect to the vector ${\bf n}$  in  Eq.(\ref{Eret_eps})  because they vanish after the integration over solid  angles  $d \Omega$ .

Since for an arbitrary constant vector ${\bf a}$ we have
\begin{equation}
\int [ {\bf n} ({\bf n} {\bf a}) + {\bf a} ] \,  \frac{d \Omega}{4 \pi}
= \frac{4}{3} \, {\bf a} ,
\label{int}
\end{equation}
 we obtain  Eq.(\ref{K_s33}) for the  electromagnetic self-force of an electron, where
\begin{equation}
m_e = \frac{1}{c^2} \, \int \!\!\! \int \! \frac{dq dq'}{2\varepsilon} = \frac{1}{c^2} \frac{q^2}{2R}
\label{m_e}
\end{equation}
is the electromagnetic mass of an electron  corresponding  to the self-energy in Eq.(\ref{Uself}). The last expression in Eq. (\ref{m_e}) is most easily obtained from the capacitor formula, $U = (1/2) q V$.  The second term on  the right hand side of Eq. (\ref{K_s33}), obtained from the last term in Eq. (\ref{K_s1}),  gives the expression in Eq.(\ref{E_charge}) for the damping force. 

Thus, we see  that  the factor $4/3$   appears in the equation of motion,  violating  the equivalence between electromagnetic mass and self-energy.   However, the calculation  above rests on the  assumption that  at a given  instant  all parts  of the electron  have the same acceleration  in a  reference frame in which they are all at rest simultaneously.  As we shall discuss below, this was the assumption challenged by Kalckar, Lindhard and Ulfbeck in \cite{Kalckar}.

\section{Acceleration of rigid body and the KLU solution}

 We define the classical electron as a rigid sphere (or spherical shell)  with small but finite extension and total charge $q$  with a spherically symmetric distribution.  We shall now show that the equivalence between electromagnetic energy and mass is restored if a proper relativistic treatment of the acceleration of a rigid body is introduced. As it turns out, this breaks the spherical  symmetry which eliminates the contribution to the self-force  from the dominant Coulomb term in Eq.(\ref{Eret_eps}).

\subsection{Relativistic description of accelerated rigid body}

Suppose that at each stage of the motion of the electron there is an inertial frame of reference in which the velocities of all the components of the electron vanish simultaneously and that  all distances between them remain unchanged in this electron rest frame while the electron moves in an arbitrary manner in the laboratory frame  $ K $ \cite{Kalckar}. This corresponds to the relativistic definition of translational motion of a rigid body, introduced by Born \cite{Born1909_v30}.
 
For simplicity we consider  one-dimensional motion of two point-like particles. The motion  occurs in such a way that the distance between them, $l_0$,  remains constant in the common rest frame  $ K '$ (see Fig. 1).  This uniquely determines the coordinates of particle 2 as functions of the coordinates of particle 1.  Let $(x_1, t_1)$, $(x_2, t_2)$ and  $(x_1', t_1')$, $(x_2', t_2')$  be the coordinates of the two particles  in reference frames  $K$ and $K'$, respectively. 
\begin{figure}
\includegraphics[scale=0.5]{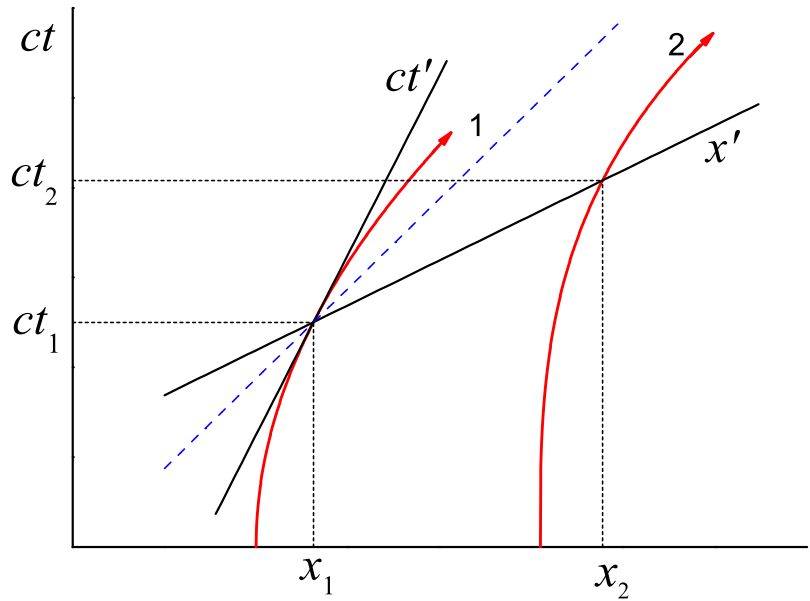}
\centering
\caption{Minkowski space-time diagram showing the motion of particles 1 and 2 in which the distance between them remains constant in the common rest frame. Both particles start out at $t=0$  with velocity zero in the laboratory system. At the event points  $(x_1, t_1)$  and  $(x_2, t_2)$  the velocities are equal, $\beta \equiv \beta_1 (t_1) =\beta_2 (t_2)$, and the two events are simultaneous in the moving rest frame (primed axes).   The dashed line indicates a branch of the light cone for the event at $(x_1, t_1)$.  }
\end{figure}
The times  $t_1'$ and  $t_2'$ are equal and  therefore the Lorentz transformation from $K$  to $K' $   yields the relation
%
\begin{equation}
t_2 - t_1 = \frac{\beta}{c} (x_2 - x_1),
\label{t2_t1}
\end{equation}
where  $c\beta = c\beta_1 (t_1) = c\beta_2 (t_2) $ is the relative  velocity of the frames $K$ and $K'$. 
On the other hand, the transformation from $K'$  to  $K$  leads to the relation
\begin{equation}
x_2 - x_1 = \gamma (x_2' - x_1')  .
\label{x2_x1}
\end{equation}

Formulas (\ref{t2_t1}) and (\ref{x2_x1}) lead to the definition of a  system  with rigid acceleration:  
\begin{equation}
x_2 = x_1 (t_1) + l_0 \gamma (t_1) ,
\label{x2}
\end{equation}
\begin{equation}
t_2 = t_1 + \frac{l_0}{c} \beta (t_1) \gamma (t_1) ,
\label{t2}
\end{equation}
where $l_0 = x_2' - x_1'$ is the distance between the particles in the rest frame.

If the velocity  is not parallel to the separation $l_0$  between the particles, the product $l_0 \beta (t_1)$  in Eq. (\ref{t2}) is replaced by the product of vectors ${\bf  l}_0 \cdot \btt (t_1)$. Below we focus as in Ch. 2 on the limit of low (non-relativistic) velocities and set $\beta=0$  corresponding to $t_1 = t_2 = 0$ in Fig. 1.  This leads to the relation
\begin{equation}
 \frac{dt_2}{dt_1}=1+  \frac{1}{c}  \,\, {\bf  l}_0 \cdot \dot { \btt }(t_1)  ,\,\,\,\, \mbox{ for}\,\,\, \beta=0.  
\label{dt_2_dt_1_a}
\end{equation}
%
We  conclude  that in the common rest frame for a system of many particles the acceleration of a particle with coordinates ${\bf r}_i$  relative to a reference particle must be given by
\begin{equation}
{\bf g}_i = \frac{{\bf g}_0}{1 + {\bf r}_i \cdot {\bf g}_0/c^2} \, ,
\label{gi}
\end{equation}
where  ${\bf g}_0$  is the acceleration of the reference particle. The choice of this particle is immaterial because  Eq.(\ref{gi}) satisfies the reciprocity relation
\begin{equation}
{\bf g}_0 = \frac{{\bf g}_i}{1 - {\bf r}_i \cdot {\bf g}_i/c^2} \, .
\label{g0}
\end{equation}

\subsection{Self-forces and self-mass of accelerated electron}

Kalckar, Lindhard and Ulfbeck calculated the electromagnetic self-force and mass of the extended electron treating  it as a system with rigid acceleration \cite{Kalckar}. Their crucial insight was that for a rigid body there is not a
simple relation between the acceleration and the total force. Consider a system of particles initially at rest and  accelerated as a rigid system with force ${\bf K}_i$  on the $i'$th particle with mass $m_i$  and acceleration ${\bf g}_i$. The (non-relativistic) equation of motion for this particle is then
\begin{equation}
m_i  {\bf g}_i = {\bf K}_i .
\label{mg_K}
\end{equation}
The total mass is $M=\sum_i m_i$  and according to the relation (\ref{gi}) for a rigid body we obtain
\begin{equation}
M {\bf g}_0  = \sum_i {\bf K}_i \, (1 +{\bf r}_i \cdot {\bf g}_0/c^2),
\label{Mg_0}
\end{equation}
where as before  ${\bf g}_0$  is the acceleration of a reference particle at  ${\bf r} = 0$. In a description of the translational motion of the rigid body as the motion of a point particle with mass $M$  the force is hence given by the right-hand side of Eq.(\ref{Mg_0}) and not by the simple sum of the forces on the parts of the body.


In the formula (\ref{K_s1}) for the self-force the correction factor in Eq. (\ref{Mg_0}) is only important for the omitted Coulomb term in Eq. (\ref{Eret_eps}). This term now gives a contribution,
\begin{equation}
\Delta {\bf K}_{self} = \int \!\! \!\! \int \! dq dq' \!\! \int \!\!  \frac {\bf n}{\varepsilon^2} \frac{ ({\bf r}'  \dot{\btt} )} {c} \frac{d\Omega}{4\pi}.  
\label{Delta_K_self}
\end{equation}
Using the symmetry between the variables $q$ and $q'$ we can rewrite this integral as 
\begin{equation}
\int \!\! \!\! \int \! dq dq' \!\! \int \!\!  \frac{d\Omega}{4\pi} \frac{ {\bf r}' -{\bf r}}  {\varepsilon^3}
\frac {({\bf r}' \dot{\btt})}{c} 
\nonumber
\end{equation}
\begin{equation}
= \int \!\! \!\! \int \! dq dq' \!\! \int \!\!  \frac{d\Omega}{4\pi} \frac{ {\bf r}' -{\bf r}}  {\varepsilon^3} 
\frac {(({\bf r}' -{\bf r}) \dot{\btt})}   {2c}
\nonumber
\end{equation}
\begin{equation}
= \int \!\! \!\! \int \! dq dq' \!\! \int \!\!  \frac{d\Omega}{4\pi} \frac{ {\bf n}({\bf n}\dot{\btt}) }  {2c\varepsilon}. 
\label{integrals_K_self}
\end{equation}
The correction (\ref{Delta_K_self}) is then seen to cancel the first term in the formula (\ref{K_s1}) and hence eliminates the factor 4/3  in Eq. (\ref{int}). The self-force in the equation of motion now has the form in Eq. (\ref{K_s33}) but without the factor 4/3, in agreement with mass-energy equivalence. 

A particularly simple example demonstrating equivalence and irrelevance of Poincar\'{e}  forces is discussed in \cite{Kalckar}. Consider two particles, both with charge $q$ and mechanical mass $m$, accelerated by a force on particle 1 in the direction towards particle 2 with a strength adjusted to keep the distance $R$ between them constant in their mutual momentary rest frame. A calculation of the self-force analogous to Eq. (\ref{K_s1}) for spherical geometry gives a self-mass equal to the result expected from equivalence (Eq. (\ref{m_e})) multiplied by a factor two. However, with the correction in Eq. (\ref{Mg_0}) this factor is reduced to unity. Obviously, there is in this example no room for introduction of Poincar\'{e} forces (see also \cite{Dillon}, \cite{Ori}). 

\subsection{Addition of forces and simultaneity}

The modification in Eq.(\ref{Mg_0}) of the relation between forces and mass for a rigid body may look very odd but,
 the modification has a simple interpretation. When a rigid body originally at rest is
accelerated during a time interval  $dt_0$  for the reference point, other points on the body are accelerated through
different time intervals, as expressed by Eq.(\ref{dt_2_dt_1_a})  and illustrated in Fig.1.
For the change of the momentum of the rigid body, defined as the sum of the momenta of its parts for fixed time in the momentary rest frame, we therefore obtain
\begin{equation}
\frac{d {\bf P}}{dt_0} = \frac{d}{dt_0} \sum_i {\bf p}_i =  \sum_i \frac{d {\bf p}_i }{dt_i} \frac{dt_i}{dt_0} =
 \sum_i {\bf K}_i (1 + {\bf r}_i  \! \cdot \!  {\bf g}_0 /c^2),
\label{dPdt0}
\end{equation}
 in agreement with the relation (\ref{Mg_0}).  As stressed in   \cite{Kalckar}, it is important that this relation is not based on any new definition but follows directly from Born's definition of a rigid body and the point dynamics of its parts. We may therefore instead regard the combination of Eqs. (\ref{Mg_0}) and (\ref{dPdt0})  as demanding the definition of the momentum given above Eq. (\ref{dPdt0}) in a description of the rigid-body motion as that of  point particle with the total mass $M$  of the body, $M=\sum_i m_i$.

Alternatively, we can describe  the acceleration process  in the accelerated reference system following the motion of the
body. Here the time differences are ascribed to different rates of clocks at different positions, i.e., to time
dilation in a gravitational field. In  \cite{Kalckar}  such a reference system is called a M\o ller box, with reference to the
book on  relativity by M\o ller \cite{Moller}. When a metric containing the spatial variation of the rate of clocks is introduced, the expression  (\ref{dPdt0}) becomes just the addition of forces on the body so the 4/3-problem disappears in a natural way. However, as noted in \cite{Kalckar}, the price to be paid for this simplification is a more complicated equation  of motion  (Eq.(A20) in \cite{Kalckar}).

\section{Equivalence from  exchange of momentum with the surrounding field}

An alternative to the calculation of self-forces is an analysis of the exchange of electromagnetic energy and momentum between the electron and the space outside. As we shall see, the 4/3-problem arises again but it can be resolved by application of a correction analogous to the KLU prescription in Eq. (\ref{Mg_0}). As mentioned, Dirac developed a relativistic analysis of the energy-momentum exchange between a region around a point electron and its surroundings in a famous paper from 1938 (see App. D). This analysis contains a solution of the 4/3-problem equivalent to that in  \cite{Kalckar}, discussed in Ch.3. However, to avoid the problem of infinite self-energy (and self-mass) for a point electron Dirac retained only the resistive self-force, corresponding to the first part of the field in Eq. (\ref{Eret1}), and replaced the infinite self-mass by a finite value through a renormalization procedure.

\subsection{Currents of field energy and momentum and conservation laws}

Consider an electron  at rest at  time    $t$,  represented by a spherical shell with radius $R$ and total charge $q$.
Electromagnetic energy and momentum balance within a spherical region of radius $\varepsilon$ surrounding the electron
is given by the equations
\begin{equation}
\frac{\partial }{\partial t} \int W dV + \int {\bf E \cdot j}\,  dV =
-  \int   {\bf S} \cdot d{\bf f} ,
\label{W_balance}
\end{equation}
\begin{equation}
\frac{\partial }{\partial t} \left( P^{(f)}_m + P^{(p)}_m  \right)
= \int  \sigma_{mn} df_n ,
\label{G_balance}
\end{equation}
where  $W=(8\pi)^{-1}({\bf E}^2+{\bf H}^2)$ is the density of electromagnetic energy,  ${\bf S} = (c/4\pi) {\bf E} \times {\bf H}$  the Poynting vector, and ${\bf j}$ the electric current density. The differential surface element $d {\bf f}$ contains a surface normal ${\bf n}$ pointing out of the sphere. The vectors ${\bf P}^{(f)}$ and ${\bf P}^{(p)}$  represent the field and
particle momenta inside the sphere with radius $\varepsilon$, where 
\begin{equation}
{\bf P}^{(f)} = \frac{1}{c^2} \int {\bf S} d V .
\label{G}
\end{equation}
The matrix  $\sigma_{mn}$  is the Maxwell stress tensor with indices referring
to the coordinates $x, y, z,$  (see Eq. (33.3) in \cite{Landau})
\begin{equation}
\sigma_{mn} = \frac{1}{4 \pi} \left[ E_{m} E_{ n} + H_{m} H_{ n} - 
\frac{1}{2} \delta_{mn} (E^2+H^2) \right],
\label{stress}
\end{equation}
where $\delta_{mn}$ is the Kronecker symbol.  We use Latin letters for indices running from 1 to 3 and the
convention that an index appearing twice is to be summed over.
The integration at the left hand side of Eq.(\ref{W_balance}) and in Eq.(\ref{G}) extends over the volume of the sphere
with radius $\varepsilon$  while  the integration at  the right hand side of Eqs. (\ref{W_balance}) and (\ref{G_balance}) is over the surface of that sphere. With the chosen sign of the stress tensor in Eq. (\ref{stress}), the right hand side of Eq. (\ref{G_balance}) gives the momentum flux into the sphere.

According to Eqs. (\ref{Eret_eps}) and (\ref{Hret_eps}) 
the leading term of the Poynting vector in the vicinity of the electron is
\begin{equation}
{\bf S} = \frac{q^2}{8\pi c \varepsilon^2} {\bf n}\times ({\bf n}\times
\ddot{\btt}) ,
\label{S}
\end{equation}
i.e., it is inversely proportional to square of the distance $\varepsilon$.  
 For terms with a higher power of  $\varepsilon$  the integrals in both Eq.(\ref{W_balance})  and Eq.(\ref{G})  vanish in the limit   $\varepsilon \rightarrow 0$. 

According to Eq.  (\ref{S}),  the Poynting vector is perpendicular to the surface normal of the sphere with radius $\varepsilon$ and hence the energy flux in Eq. (\ref{W_balance}) equals zero.
The second term at the left hand side of Eq. (\ref{W_balance}) is also zero since ${\bf j}=0$  in the rest system. The field energy contained in the volume is therefore constant,
\begin{equation}
\frac{\partial }{\partial t} \int W dV = 0 \,\, ,
\label{W_zero}
\end{equation}
and we may conclude  that in its rest frame  the electron does not emit energy but only momentum, in contrast to the conclusion in \cite{Landau} based on the Larmor formula and the symmetry of the emitted radiation.  

For $\varepsilon \rightarrow R_{+}$,  the derivative of the electromagnetic field momentum  inside the sphere with radius $\varepsilon$  tends  to zero since the volume of this sphere approaches the volume inside the uniformly charged
spherical shell  where there is no field, ${\bf E} =0$.    
This implies  that, in this limit, the right hand side of Eq.(\ref{G_balance}) represents the rate of change of mechanical momentum, only, i.e. it represents the self-force,
\begin{equation}
{\bf K}_{self}=\frac{d {\bf P}}{dt}=\int \!\!\! \int \! {\bf k}_s df \, \, .
\label{K_s3}
\end{equation}
From Eq.  (\ref{stress}) we have 
\begin{equation}
{\bf k}_s=\frac{1}{4\pi}\left[{\bf E}({\bf E}{\bf n}_s) -
\frac{1}{2}{\bf n}_s |{\bf E}|^2 \right] ,
\label{k}
\end{equation}
where we have introduced an index $s$ on the surface normal and on the momentum  flux at the surface of integration.
The magnetic-field terms in Eq.(\ref{stress}) are of second order in $ \varepsilon$ after the integration (\ref{K_s3}) and can be ignored.

\subsection{Self-force from transport of field momentum}

Here we treat the simple  case with $R\ll \varepsilon$, the point electron.  As shown in App. B,
the result applies for all $\varepsilon > R$.   
With the approximation above the charge is located at the center of the sphere so the unit vector ${\bf n}$ in Eq. (\ref{Eret_eps}) is equal to the surface normal ${\bf n}_s$.
Inserting the expression (\ref{Eret_eps})  for the electric field into Eq.(\ref{k}) and keeping only terms of order $ \varepsilon^{-2}$  or lower we obtain
for the momentum flux into the sphere with radius $\varepsilon$
$$
{\bf k}_s = \frac{q^2}{8\pi \varepsilon^2 } \, \left\{ \frac{{\bf n}_s} {\varepsilon^2} -
\frac{1}{c\varepsilon } \left[  {\bf n}_s  ({\bf n}_s  \dot{\btt}) +
\dot{\btt} \right]  \right. + 
$$
\begin{equation}
\left.
+ \frac{1}{c^2} \left[ \, ({\bf n}_s \dot{\btt})^2 {\bf n}_s 
+ \frac{5}{2} \, ({\bf n}_s  \dot{\btt}) \dot{\btt} -  |\dot{\btt}|^2
{\bf n}_s +  \frac{4}{3} \, \ddot{\btt} \, \right] \right\}\,\,\, . 
\label{k_eps}
\end{equation}
The first (Coulomb) term and the first three terms in the last parenthesis give no contribution to the integral in
Eq.(\ref{K_s3}) owing to their odd symmetry under change of sign of ${\bf n}_s$. Using the relation (\ref{int}) we then obtain for the force on the system consisting of the charged sphere and the field  inside the distance $\varepsilon$
\begin{equation}
{\bf K}_{s} (\varepsilon) \approx  - \frac{4}{3}  \frac{q^2}{2c\varepsilon} \dot{\btt} + \frac{2}{3} \frac{q^2}{c^2}
 \ddot{\btt} .
\label{K_s2}
\end{equation}
This agrees with Eq.(\ref{K_s33}) with $m_e$  given by Eq.(\ref{m_e}),  except for the replacement of $R$ by $\varepsilon$.
This difference is consistent with the notion that the electromagnetic energy is located in the field with the density given below Eq.(\ref{G_balance}). The dominant term is the Coulomb field and  the energy inside a spherical  shell with volume 
 $4 \pi r^2 dr$  is proportional to $ r^{-2}$. The total field energy outside the distance  $\varepsilon$  is therefore proportional to $\varepsilon^{-1}$   and  for $\varepsilon =R$ it equals the energy  $q^2/2R$   of a uniformly charged sphere. The negative of the first term in Eq.(\ref{K_s2}) is the force required to give this field the acceleration  $c \dot{\btt}$   (apart from the troublesome factor 4/3 which we discuss below).

According to the KLU prescription, the factor 4/3 in this formula may be eliminated by introduction of the
relativistic correction factor  $(1+{\bf r}_i \cdot \dot{\btt} /c )$   in the sum over forces ${\bf K}_i$ acting at ${\bf r}_i$  on a rigid body with acceleration  $c \dot{\btt}$   at ${\bf  r} = 0$. We should include this factor in the integration since the sphere is defined in the electron's rest frame and follows its motion as a rigid body. 
For the point electron the momentum flux is given by Eq.(\ref{k_eps}). The correction factor is only important for the Coulomb term and we obtain
$$
 \Delta {\bf K}_{s} (\varepsilon) =
- \Delta m_e c \dot{\btt} = \frac{q^2}{8\pi \varepsilon^2} \int d\Omega_s 
\left( 1+ ({\bf n}_s   \dot{\btt}) \varepsilon / c \right) {\bf n}_s  = 
$$
\begin{equation}
= \frac{q^2}{4c\varepsilon} \int_{-1}^1 d\cos \theta \cos^2  \theta \dot{\btt} = \frac{q^2}{6c\varepsilon} \dot{\btt} .
\label{integration_4_3}
\end{equation}
With this correction to Eq.(\ref{K_s2})  we obtain full relativistic equivalence between mass and energy of  the field outside the sphere with radius  $\varepsilon$.  In the limit  $\varepsilon \rightarrow R$  we obtain equivalence between the total field energy and mass. The calculation gives a physical interpretation of the electron's self-force as the drag by the inertial mass of the electromagnetic field. 

\section{Energy-momentum tensor}

The KLU paper clearly identified the error in the standard calculation of the electromagnetic self-mass of
an electron represented by a classical model of a rigid, charged spherical shell. However, it remains to be explained what exactly is wrong with the definition in Eq. (\ref{P_el}) of the momentum of the field associated with
the electron and with Eqs.(\ref{W_balance}), (\ref{G_balance}), expressing conservation of the total energy and momentum of particles and fields. Is the redefinition of the field momentum and energy suggested by Rohrlich  correct and, if so, what is the justification? In this chapter we elucidate these questions based on the general discussion of field energy-momentum in \cite{Landau} \S 32 and the application to the electromagnetic field in \S 33.

We shall use standard four-dimensional relativistic notation, with Greek indices running from 0 to 3 for four-vectors transforming like the time-space coordinates of an event, $x^{\mu} = (ct, x, y, z)$. In addition to these
(contravariant) vectors we introduce the corresponding vectors with opposite sign of the last three components, called covariant vectors and distinguished by lower indices, $x_{\mu}$. The invariant scalar product of two four-vectors $x^{\mu}$ and $y^{\mu}$ can then be written as $x^{\mu}y_{\mu}$  with the convention of summing over indices
appearing twice. Tensors of rank two, $A^{\mu \nu}$, transform like the product of the components of two four-vectors. As for four-vectors, the indices can be moved up or down with the convention that this changes the sign for the spatial indices 1,2,3 but not for the time index 0.

\subsection{Electromagnetic field tensor}
The electromagnetic potentials may be combined into a four vector $A^{\mu} = (\varphi, {\bf A})$ and the sources of the fields into a four-current $j^{\mu} = (c\rho, {\bf j})$. A compact representation of the fields is the electromagnetic field tensor defined by \cite{Landau}
\begin{equation}
F_{\mu \nu} = \frac{\partial A_\mu}{\partial x^\nu}  - \frac{\partial A_\nu}{\partial x^\mu} .
\label{F_mn}
\end{equation}
The tensor is antisymmetric in the indices $\mu, \nu$. The time derivative of the kinetic energy and momentum of a particle with rest mass $m$ and charge $q$, interacting with the fields through the Lorentz force, are then determined by the equation of motion,
\begin{equation}
mc \frac{d u_\mu}{ds } =    \frac{q}{c} F_{\mu \nu} u^\nu ,
\label{Eq_motion_in_F}
\end{equation}
where $u^\mu =(\gamma , \gamma \btt)$ is the four-velocity and $ds=cdt/\gamma$. 

Also the Maxwell equations for the fields can be written in a compact form. 
The equations without source terms can be expressed as the following relation for the field tensor,
\begin{equation}
\frac {\partial F_{\mu \nu}}{\partial x^\xi}  + \frac {\partial F_{\nu \xi}}{\partial x^\mu} +
 \frac {\partial F_{\xi \mu }}{\partial x^\nu} =0,
\label{4_Max_eq_1}
\end{equation}
and the equations relating the fields to the sources as the relation
\begin{equation}
\frac {\partial F^{\mu \nu}}{\partial x^\nu} =   - \frac {4 \pi }{c } j^\mu  .
\label{4_Max_eq_2}
\end{equation}

\subsection{Energy-momentum four-vector of field around moving electron}

The energy and momentum densities of the electromagnetic field may be expressed through an energy-momentum tensor 
 (Eq. (32.15) in \cite{Landau}),
\begin{eqnarray}
T^{\alpha \beta} =
\left( \begin{array}{cccc}
W              & S_{x} / c         &   S_{y} / c        & S_{z} / c       \\
S_{x} / c   & - \sigma_{xx}   & -\sigma_{xy}   &  -\sigma_{xz} \\
S_y / c      &  - \sigma_{yx}    &  -\sigma_{yy} &  -\sigma_{yz} \\
S_z / c       & -\sigma_{zx}    &  -\sigma_{zy}   &  -\sigma_{zz} 
\end{array} \right).
\label{Tensor_T}
\end{eqnarray}
Here $W$ is the energy density,  ${\bf S}$  the Poynting vector, and  $\sigma_{mn}$  the Maxwell stress tensor, given explicitly in Eq. (\ref{stress}). The energy-momentum tensor may also be expressed in terms of the field tensor (Eq. (33.1) in  \cite{Landau})
\begin{equation}
{T^\mu}_\nu = \frac {1}{4 \pi} \left(   -  F_{\nu \xi}  F^{\mu \xi}  + \frac{1}{4} \delta_{\mu \nu} 
 F_{\xi \eta}  F^{\xi \eta} \right), 
\label{EM_4_tensor}
\end{equation}
where $ \delta_{\mu \nu}$ is the Kronecker symbol. 

The energy and momentum of the field on a hyperplane in 4-dimensional space is then given by
\begin{equation}
P^{\alpha} = \frac{1}{c} \int T^{\alpha \beta}dS_{\beta},
\label{4_momentum}
\end{equation}
where the differential is an element of the hyperplane, $d\sigma$,  multiplied by a time-like unit four-vector  perpendicular to that plane and in the future light cone, $dS_{\beta}=n_{\beta} d \sigma$. For the special case of a hyperplane defined by $x^0 = const.$, corresponding to the 3-dimensional space in the laboratory frame $K$,  we obtain the standard expression for  the energy as an integral over space of the energy density $W$ and of   the momentum as an integral of the Poynting vector divided by $c^2$.   As discussed in Ch. 2 below  Eq. (\ref{P_el})  this expression for the energy-momentum of the field fails to transform as a four-vector.

On the other hand, with the choice $n_\beta = u_\beta$  the hyperplane corresponds to the 3-dimensional space in the electron rest frame $K'$ and Eq.(\ref{4_momentum}) is  Rohrlich's alternative definition of the energy-momentum vector of the electromagnetic field associated with the electron. In the rest frame we have $u_\beta = (1,0,0,0)$  and Rohrlich's formula gives the same result as the standard one, discussed in Eq. (\ref{P_el}) and below. However, since 
  $G^\alpha = T^{\alpha \beta} u_\beta$  is a four-vector and the surface element $d\sigma$ in the rest frame is an invariant, the energy-momentum vector defined in Eq.(\ref{4_momentum}) transforms  as a four-vector and the 4/3-problem should disappear. As an illustration of the content of formula (\ref{4_momentum})   we verify this by direct calculation.

In Eq.(\ref{4_momentum})  the integration is over space in the rest frame $K'$  while the energy-momentum tensor is defined in the laboratory, so we first express $G^{\alpha}$ in the primed coordinates of the rest frame. We assume that the velocity $c\btt$  is in the $x$-direction and obtain from the Lorentz transformation 
${\bf r}' = (x',y',z')=(\gamma (x-\beta ct),y,z)$. The electromagnetic field from a charge $q$  in uniform motion is given by (\cite{Landau}  Eq.(38.6))
\begin{equation}
{\bf E}({\bf r},t)=  q \gamma  \frac{ (x-\beta ct,y,z) } {\left[ \gamma^2(x-\beta ct)^2+y^2+z^2 \right]^{3/2}}, \,\,\,\,
{\bf H}=\btt \times {\bf E}.
\label{E_uniform_motion}
\end{equation}
Expressed as a function of the coordinates in $ K'$   the electric field is given by
\begin{equation}
{\bf E}({\bf r}') = q \gamma \frac { (x'/\gamma, y',z')}{r'^3}.
\label{E_rest_Coulomb}
\end{equation}
We then calculate the spatial part of $G^\alpha$,    ${\bf G}=(G_x,G_y,G_z)$, 
$$
G_x=\frac{\gamma}{c} S_x +\gamma \beta \sigma_{xx}=
$$
$$
= \frac{\gamma \beta}{4 \pi} \left( E^2 - E_x^2  \right) +
\frac{\gamma \beta}{8 \pi} \left( 2E_x^2 -\left(E^2 +\beta^2 \left[ E_y^2 + E_z^2  \right]  \right)  \right) =
$$
\begin{equation}
= \frac{ \beta}{8 \pi \gamma} \left(E^2 +\beta^2 \gamma^2  E_x^2 \right),
\label{G1}
\end{equation}
$$
G_y=\frac{\gamma}{c} S_y +\gamma \beta \sigma_{yx}=0,  \,\,\, \, \,\,\, G_z=\frac{\gamma}{c} S_z +\gamma \beta \sigma_{zx}=0.
$$

The field is zero inside the sphere with radius $R$ and hence the field momentum is given by the integral
\begin{equation}
P_x  = \gamma \frac {q^2 \beta }{8\pi c}   \int_R^\infty     \frac{4\pi r'^2 dr'} {r'^4} =
\gamma \left( \frac{1}{c^2} \frac{q^2}{2R} \right) \beta c = \gamma m_e \beta c  
\label{Px_uniform}
\end{equation}
and the result is consistent with the relativistic relation between mass and energy in Eq.(\ref{m_e}).

\subsection{Mechanical energy-momentum tensor}

To justify the new definition of the field energy and momentum it is necessary  to verify that it is consistent with the exchange of energy and momentum between particles and field, as determined by the Maxwell equations and the Lorentz force. For this purpose we introduce an analogous mechanical energy-momentum tensor for particles with rest-mass density
\begin{equation}
\mu = \sum_j  \mu_j =  \sum_j  m_j \delta ( {\bf r} - {\bf r}_j),
\label{mass_density}
\end{equation}
where ${\bf r}_j $  is the position vector of the mass $m_{j}$. The energy-momentum tensor becomes (\cite{Landau} Eq. (33.5))
\begin{equation}
{T^{(p)}}^{\alpha \beta} = \sum_j  \mu_j  c \frac{dx_j^{\alpha}}{ds_j}  \frac{dx_j^{\beta}}{dt} =
\sum_j   \mu_j  c u_j^{\alpha} u_j^{\beta} \frac{ds_j}{dt}.
\label{mass_tensor}
\end{equation}
Applying Eq.(\ref{4_momentum}) for the hyperplane with $x^0 = const.$  we obtain the required result,
\begin{equation}
P^{\alpha} = \int  \sum_j \mu_j  c u_j^{\alpha} dV   = \sum_j  m_j  c u_j^{\alpha}.
\label{mass_momentum}
\end{equation}

As discussed in Ch. 4, below Eq.(\ref{dPdt0}), the momentum of a rigid body is the sum of the momenta of its parts for fixed time $t'$  in the rest system $K'$.  This corresponds to integration in Eq.(\ref{4_momentum}) over the hyperplane perpendicular to the velocity  $u^\beta$  of the rest frame of the body with $d\sigma = d^3 {\bf r'}$, leading to
\begin{equation}
P^\alpha = \frac{c}{\gamma} \int d^3 {\bf r'} \sum_j  m_j \delta \left( {\bf r} - {\bf r}_j   \right)  u_j^\alpha , 
\label{mass_momentum1}
\end{equation}
where the coordinates ${\bf r'}_j$   indicate the positions of the mass elements $m_j$   and  $u_j^\alpha = u^\alpha $ the velocities   for fixed $t'$.  As in the calculation
above of the field momentum, we assume that the velocity is in the $x$-direction and introduce the primed  variables, 
$ {\bf r'} = \left(  \gamma (x - c\beta t), y, z  \right)$, in the $\delta$-function with the replacement 
$ {\bf r} \rightarrow  \left(  \frac{\textstyle{ x'}}{ \textstyle{ \gamma} }  +  c\beta t, y', z'  \right)$.   
The positions of the mass elements move with the speed  $c\beta$  in the $x$-direction and the two terms  proportional to $t$   in the $\delta$-function cancel. The integration then gives a factor $\gamma$  and we again obtain the result in Eq.(\ref{mass_momentum})  but with the velocities for fixed $t'$ so that we obtain the simple relation for a point particle with velocity $u^\alpha$
\begin{equation}
P^\alpha   = \sum_j  m_j  c u_j^{\alpha} = Mcu^\alpha , 
\label{mass_momentum2}
\end{equation}
where $M=\sum_j m_j$ is the total rest mass of the body.  


\subsection{Conservation of total energy and momentum of particles and fields}

The change with time of the energy and momentum is related through Gauss'  theorem to the four-divergence of the energy-momentum tensor.
For a system of charged particles interacting with the electromagnetic field the total energy-momentum tensor is the sum of the particle and field tensors,  ${T^{\mu}}_{\nu} = {{T^{(p)}}^{\mu}}_{\nu} + {{T^{(f)}}^{\mu}}_{\nu}$.  
With both charges and fields in the volume, the particle and field tensors are not separately divergence free but, as demonstrated in   \cite{Landau} \S 33,  the total energy-momentum tensor is,
\begin{equation}
\frac{\partial}{\partial x^{\mu}}  \left( {{T^{(p)}}^{\mu}}_{\nu} + {{T^{(f)}}^{\mu}}_{\nu}  \right) = 0.
\label{total_em_tensor}
\end{equation}

To show this we differentiate Eq. (\ref{EM_4_tensor}) and obtain 
\begin{equation}
\frac{ \partial {T^{(f) \mu}}_\nu  } {\partial x^\mu}  = \frac {1}{4 \pi}  \left(   \frac{1}{2} 
 F^{\xi \eta}  \frac{ \partial F_{ \xi \eta }} {\partial x^\nu}  -
 \frac{ \partial F_{ \nu \xi }} {\partial x^\mu}  F^{\mu \xi}   -  
 F_{\nu \xi}    \frac{ \partial F^{ \mu \xi }} {\partial x^\mu}  \right) .  
\label{differential_EM_4_tensor}
\end{equation}
We replace the last factor in the first term using the relation (\ref{4_Max_eq_1})  and the second factor in the last term   using the relation (\ref{4_Max_eq_2}).  This leads to 
\begin{equation}
\frac{ \partial {T^{(f) \mu}}_\nu  } {\partial x^\mu}  = \frac {1}{4 \pi}  \left(   
 - \frac{1}{2}  F^{\xi \eta}  \frac{ \partial F_{  \eta \nu }} {\partial x^\xi} 
- \frac{1}{2}  F^{\xi \eta}  \frac{ \partial F_{  \nu \xi }} {\partial x^\eta} -
\right.
\nonumber
\end{equation}
\begin{equation}  
\left.
   -  F^{\mu \xi}  \frac{ \partial F_{ \nu \xi }} {\partial x^\mu}    -  
\frac{4\pi}{c}    F_{\nu \xi} j^\xi   \right) .  
\label{differential_EM_4_tensor_1}
\end{equation}
By renaming the indices one can easily verify that the third term cancels the first two. This leaves the result
\begin{equation}
\frac{ \partial {T^{(f) \mu}}_\nu  } {\partial x^\mu}  = - \frac{1}{c}    F_{\nu \xi} j^\xi .
\label{differential_EM_4_tensor_2}
\end{equation}
Next we consider the energy-momentum tensor for the particles in Eq. (\ref{mass_tensor}). 
First we assume that all particles have the same velocity. The four-divergence of the tensor then becomes
\begin{equation}
\frac{ \partial {T^{(p) \xi}}_\nu  } {\partial x^\xi}  = c u_\nu  \frac{\partial}{\partial x^\xi} \left( \mu \frac{d x^\xi}{dt} \right) +
\mu c \frac{d x^\xi}{dt} \frac{\partial }{\partial x^\xi} u_\nu.
\label{differential_mass_4_tensor}
\end{equation}
If we replace $\mu$ by a continuous rest-mass distribution the first term is proportional to the four-divergence of the mass current which is zero due to conservation of rest mass. In the second term, we introduce the equation of motion in
 Eq. (\ref{Eq_motion_in_F}) which for a continuous charge distribution $\rho$ may be written as
\begin{equation}
\mu c \frac{d u_\nu}{dt } =   \frac{c}{\gamma}  \frac{\rho}{c} F_{\nu \xi} u^\xi  =\frac{1}{c}  F_{\nu \xi} j^\xi .
\label{Eq_motion_in_F_2}
\end{equation}
If the particles, or mass elements, do not have the same velocity, as is the case for an accelerated rigid body viewed from the laboratory frame (Fig. 1), the result in Eq. (\ref{Eq_motion_in_F_2}) is obtained for each small mass element and added together they again give the negative of Eq. (\ref{differential_EM_4_tensor_2}).
 Combining with Eq. (\ref{differential_EM_4_tensor_2}) we obtain the desired result in Eq. (\ref{total_em_tensor}).

Integration of Eq. (\ref{total_em_tensor}) over a region of four-space, delimited by two hyperplanes with constant times  $t$ and $t+dt$ and a spherical surface $f$, and application of Gauss'   theorem leads to the equations   (\ref{W_balance}) and (\ref{G_balance})  which were the starting point for our calculations in Ch.4.
However, we can now also see the problem with these relations. The integrals for fixed times $t$  and $t+dt$  of the energy-momentum tensor for the particles, i.e. for the components of the rigid body, do not represent the energy and momentum we associate with the rigid body. 
If we want to represent the motion of the body as that of a point mass, these quantities should be calculated on hyperplanes corresponding to fixed times in the momentary rest frames of the rigid body. In order to apply Gauss' theorem to Eq.(\ref{total_em_tensor}) we must then also integrate the energy-momentum tensor of the field over 
a volume delimited by these hyperplanes. 

The expression for the momentum four-vector in Eq.(\ref{4_momentum})  with $n_\beta = u_\beta$    is identical to the one suggested in  \cite{Rohrlich1990} but the justification is different. The modification of the standard expression   is imposed by the relativistic definition of a rigid body introduced by Born and corrected in \cite{Kalckar}. 
For consistency of the description  the same hyperplane must be chosen for definition of the energy and momentum of the field as  for the rigid body. Thus the justification is not just a requirement of covariance and there is  not the freedom of definition  implied by Jackson, \cite{Jackson} Ch.16. The field energy-momentum defined in Eq. (\ref{4_momentum}) would be covariant with any choice of a fixed hyperplane for the integration. 
 Furthermore, the introduction of Poincar\'e stresses to solve the 4/3-problem is not only unnecessary but is hiding the real origin of the problem. 

\section{Summary and concluding remarks}

The linked problems in classical electrodynamics of the electromagnetic mass of an electron and the damping
of its motion due to emission of radiation have a long and interesting history. 
Early work on a classical model of the electron, in particular by Abraham \cite{Abraham} and Lorentz \cite{Lorentz}, focused on the damping and led to the Abraham-Lorentz equation of motion, as discussed in \cite{Jackson} Ch.16. 
Calculation of the electromagnetic mass required a description of the motion of a rigid body, and Born is credited with being the first to formulate the relativistic concept of a rigid body in a series of papers published around 1910. 
In the description of the motion of a body by a bundle of trajectories in 4-dimensional space its shape in the momentary rest frame is determined as the cut of this bundle with a  3-dimensional hyperplane perpendicular to the four-velocity, and rigidity requires this shape to be conserved \cite{Born1909_v30}. 

With this definition Born considered the Abraham-Lo\-rentz model of an electron as a rigid, uniformly charged spherical shell and calculated the self-force and the corresponding electromagnetic mass as $m=(4/3) U_{el}/c^2$,  where $U_{el}$  is the electrostatic energy, in violation of the principle of equivalence between mass and energy in the theory of special relativity,  expressed in Einstein's famous equation, $E=mc^2$. The crucial mistake in this calculation was Born's failure to realize the full consequences of relativity: ``We will understand as the resulting force of a force field, the integral of the product of rest charge and rest force [field] over the rest shape of the electron''  \cite{Born1909_v30} Ch.3, \S 11. The problem with this seemingly  innocuous definition was not realized at the time  and instead a remedy was suggested by Poincar\' e \cite{Poincare}. In the Abraham-Lorentz model of the electron additional forces, so-called Poincar\' e stresses, are required for stability, and the combined contributions from these and the electromagnetic forces to the mass and energy of the electron could be in accordance with Einstein's principle of equivalence \cite{Jackson} Ch.16.

As shown by Kalckar, Lindhard and Ulfbeck  \cite{Kalckar} and discussed here in Ch.  3, the conflict with
relativistic equivalence is resolved when the relativistic modification of Born's  definition of total force is taken into
account. 
An alternative to a calculation of internal forces between charge elements is an evaluation of the energy-momentum
transport through a surface surrounding the charged spherical shell.  As demonstrated in Ch. 4, the result of this type of calculation is consistent with equivalence between energy and inertial mass of the field when the time differences in rigid acceleration of the field are taken into account. This was seen to hold not only for the total field outside the spherical charged shell but for the field outside a sphere with arbitrary radius. In this sense, we have demonstrated detailed equivalence between mass and energy for the electromagnetic field around an accelerated electron.
The question of equivalence for an atomic system was discussed in \cite{Kalckar} and it was shown that it is independent of whether a classical description is used or a quantal description like the Dirac equation.

Curiously, a solution of the 4/3-problem was suggested already around 1920 by Enrico Fermi \cite{Fermi1922}, as we have  discussed in App. C.
This paper and other related early Fermi papers have recently been reviewed and extended 
by Jantzen and Ruffini  \cite{Jantzen2012}.  Fermi did not clearly identify the problem with Born's definition of total force on a rigid body, revealed in \cite{Kalckar}, but pointed in the right direction for solution of the 4/3-paradox. His suggestion was largely ignored and forgotten  at the time but was taken up by Rohrlich \cite{Rohrlich1990}, who suggested adoption of the new, covariant definition of field momentum derived by Fermi. 
 ``However, this will hardly do''  was the brief comment in  \cite{Kalckar}. One cannot arbitrarily redefine the field momentum. It must be demonstrated that the definition is consistent with the exchange of energy and momentum between the particles and fields. This we have done  in Ch.5. 

Thus we have arrived at a comprehensive solution of the 4/3 paradox: in a description of the motion of a charged, rigid sphere by the dynamics of a point charge with the total mass of the body, the energy and momentum must be evaluated as a sum over the elements of the body for fixed time in its momentary rest frame. The factor 4/3 then disappears from the electromagnetic self-mass obtained from the self-force on an accelerated body. For consistency of the description, the energy and momentum of the electromagnetic field associated with the charge must then  also be evaluated for fixed time in the momentary rest frame of the body. The energy-momentum vectors for the particle and the field, calculated in different reference frames, then refer to the same physical quantities, i.e., they are evaluated as sums over the same event points, and hence they transform as 4-vectors \cite{Gamba}.

 The authors are  especially indebted  to the  late professor Jens Lindhard for numerous discussions and criticism on this issue in the first half of the 1990s. The authors are also grateful to E.Bonderup for detailed constructive criticism and to A.Kh.Khokonov for useful discussions and interest in this work.

\section*{Appendix A.  Expansion of electromagnetic fields near accelerated charge}

The retarded electromagnetic field at a space-time point $({\bf r}, t)$, produced by a point charge $q$ carrying out an
assigned motion ${\bf r}_0 (t')$, is determined by the state of motion of the charge at an earlier time  $t'$  (see \cite{Jackson} formulas (14.13) and (14.14), or  \cite{Landau} formula  (63.8)),
$$
{\bf E}_{ret} ({\bf r}, t) = \frac {q}{R'^2} \frac{(1-\beta '^2) ({\bf n}' -
\btt')}{(1-{\bf n}' \!\! \cdot \! \btt')^3} + \frac  {q}{c R'}
\frac {{\bf n}' \times [ ({\bf n}' - \btt') \times \dot{\btt}']}
{(1-{\bf n}' \!\! \cdot \! \btt')^3},
\eqno(\mbox{A}1)
$$
$$
{\bf H}_{ret} ({\bf r}, t) = {\bf n}' \times {\bf E}_{ret} ({\bf r}, t), 
\eqno(\mbox{A}2)
$$
where  $\btt' \equiv \btt (t') = {\bf v}(t')/c$  is the velocity of the charge (relative to the speed of light) and  $\dot{\btt}'=d \btt' /dt'$ is the acceleration (divided by $c$), ${\bf n}' \equiv {\bf n} (t')$  is a unit vector in the direction towards the  observation point,  $\bf{r}$,  from the electron position at time   $t'$  (i.e., in the direction of ${\bf r} -  {\bf r}_0 (t')$ )  and   $R' \equiv R (t')$ = $\mid \! {\bf R} (t') \! \mid \equiv \mid \!  {\bf r} -  {\bf r}_0 (t') \! \mid$. 
The primed  quantities refer to the  time  $t'$ defined as  
$$
t' = t - \frac{1}{c} \mid {\bf r} -  {\bf r}_0 (t') \mid .
\eqno(\mbox{A}3)
$$
The expression for the advanced field,  ${\bf E}_{adv}$, can be obtained from Eq.(A1) by a change of variables:  $\btt' \rightarrow - \btt'$ and modification of Eq.(A3) to  $t' = t + R'/c$. After the following expansions, leading to formulas expressed in variables related to the electron motion at time $t$, the corresponding formulas for the advanced field are obtained  simply by a  change of sign of the velocity and its second derivative. 
\begin{figure}
\includegraphics[ width=1.0\textwidth]{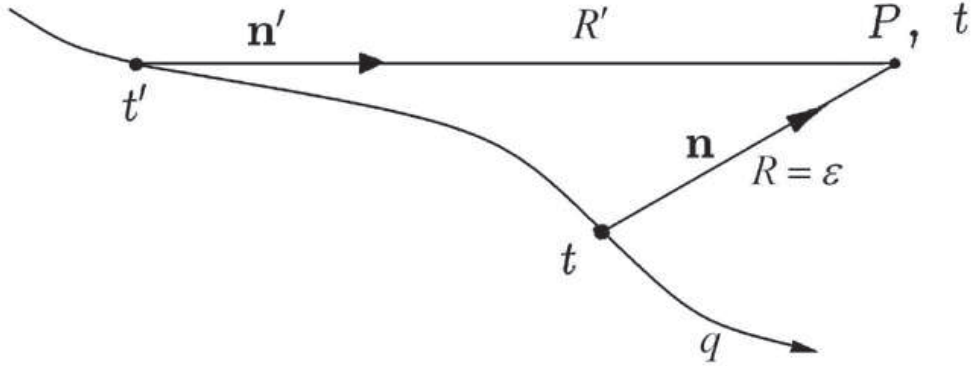}
\centering
\caption{ Illustration of the geometry for calculation of the retarded field in the vicinity of an electron.}
\nonumber
\end{figure}

We are interested in the retarded  electric field  (A1) in the  neighborhood of a moving charge $q$. 
The notation for the calculation of the field at time $t$ at a point $P$ with coordinate vector $\bf{r}$ is illustrated in Fig.2.
Our aim is to express the field as a function  of the radius vector of the point relative to the  position of the charge at the same time $t$, $  {\bf R} (t) =  {\bf r} - {\bf r}_0 (t) \equiv   {\bf n} \varepsilon$,  the acceleration, $c \dot{\btt} (t)$,   and the derivative of the acceleration, $c \ddot{\btt} (t)$. For simplicity, we perform the calculations in the rest frame of the charge at time $t$,
$$
\btt (t) =0 \, .
\eqno(\mbox{A}4)
$$
We consider a variation of the distance $\varepsilon$ of the point $P$ from the charge for fixed ${\bf n}$.  The position of the charge at time $t$ is fixed but the position at the earlier time   $t' =t - R'/c $ is a function of $\varepsilon$ through the delay $\tau \equiv R' /c$. We want to expand the field (A1) in the parameter $\varepsilon$.  The vector       ${\bf R}(t) - {\bf R} (t') =  {\bf r}_0 (t') -  {\bf r}_0 (t) $ depends on  $\varepsilon $ only through the parameter $\tau$ and we may therefore first expand this vector in $\tau$ and  we obtain
$$
{\bf R}(t')  \approx
{\bf R}(t) - \tau \dot{\bf R} (t)  + \frac{1}{2} \, \tau^2 \ddot{\bf R} (t) 
- \frac{1}{6} \, \tau^3 \stackrel{...}{\bf R} (t)  =
\nonumber
$$
$$
= {\bf R} - \frac{1}{2c}\, \dot{\btt} R'^2 + \frac{1}{6c^2} \,
\ddot{\btt} R'^3 .
\eqno(\mbox{A}5)
$$
To convert this expression into an expansion of $R' = \mid \! {\bf R}(t') \! \mid$ in $\varepsilon$ we note that  since $\beta = 0$  also the dependence on   $\varepsilon $  of the vector ${\bf R}(t) - {\bf R} (t') $  must be of second and higher order. Hence the series expansion of $R'$ has the form 
$$
R'  \approx \varepsilon + a_1 \varepsilon^2 + a_2 \varepsilon^3 \, \, .
\eqno(\mbox{A}6)
$$
Inserting this into Eq.(A5) and keeping terms of up to third order in $\varepsilon$  in the norm of the vector  ${\bf R} (t')$  we obtain for the coefficients  $a_1$  and $a_2$  in Eq.(A6) 
$$
a_1 = - \frac{1}{2c} \  (\dot{\btt} {\bf n})
\eqno(\mbox{A}7)
$$
$$
a_2 = \frac{3}{8c^2} \  (\dot{\btt} {\bf n})^2 + \frac{1}{6c^2} \  (\ddot{\btt} {\bf n}) +
\frac{1}{8c^2} \ |\dot{\btt}|^2 \, \, ,
\eqno(\mbox{A}8)
$$
where all quantities on the right hand side are taken at  the time  $t$.  Also  the velocity $c \btt (t')$  is a function of $\varepsilon$ only through $\tau$ and   may first be expanded in this parameter. To second order in  $\varepsilon$  this leads to
$$
\btt (t') \approx  - \frac{\varepsilon}{c} \dot{\btt} + \frac{\varepsilon^2}{2c^2} \left( ({\bf n} \dot{\btt}) \dot{\btt} + \ddot{\btt} \right) .
\eqno(\mbox{A}9)
$$
For the derivative, we need only include the first-order term $\dot{\btt}(t')  \!\! \approx \!\!  \dot{\btt} - (\varepsilon /c )  \ddot{\btt} $.

The expansion of the unit vector ${\bf n} (t')$  may be obtained from the ratio of the expressions in Eqs.(A5) and  (A6), and to second order we obtain
$$ {\bf n} (t') \approx  {\bf n} + \frac{\varepsilon}{2c} \left( ({\bf n} \dot{\btt}) {\bf n} - \dot{\btt} \right) -  $$
$$
- \frac{\varepsilon^2}{2c^2} \left(   \frac{1}{4} ({\bf n} \dot{\btt})^2 {\bf n}  + \frac{1}{3} ({\bf n} \ddot{\btt}) {\bf n} + \frac{1}{4}  |\dot{\btt}|^2 {\bf n} - \frac{1}{2}    ({\bf n} \dot{\btt}) \dot{\btt} - \frac{1}{3} \ddot{\btt} \right). 
\eqno(\mbox{A}10)
$$
Inserting these expansions into Eq. (A1) we obtain 
$$
{\bf E}_{ret} \approx \frac{q}{\varepsilon^2 } \, {\bf n} -
\frac{q}{2c\varepsilon } \left[  {\bf n} ({\bf n} \dot{\btt}) +
\dot{\btt} \right] +
$$
$$
+ \frac{q}{c^2} \left[ \, \frac{3}{8} \, ({\bf n} \dot{\btt})^2 {\bf n} 
+ \frac{3}{4} \, ({\bf n} \dot{\btt}) \dot{\btt} - \frac{3}{8} \, |\dot{\btt}|^2
{\bf n} +  \frac{2}{3} \, \ddot{\btt} \, \right] \,\,\, , 
\eqno(\mbox{A}11)
$$
and from insertion of Eqs. (A10) and (A11) into Eq. (A2),
$$
{\bf H}_{ret} \approx \frac{q}{2c^2 } \, {\bf n} \times \ddot{\btt} \,\,\, .
 \eqno(\mbox{A}12)  
$$

Expansions of this type  were first performed by  Page (see formulas (21) - (24) in \cite{Page2}).  Dirac  did the same calculations in covariant  form \cite{Dirac} and for $\beta=0$ formula (A11) is the same as the expression (60) in  \cite{Dirac}. 
Heitler also considered   the expansion (A11), retaining   terms  of even order in $\bf{n}$  only (see Eq.(14) in \cite{Heitler} \S 4).   The first two terms in Eq.(A11)  were applied in \cite{Kalckar} and characterized as a first-order expansion in the acceleration.


\section*{Appendix B.  Equivalence from flux of field momentum }

To prove complete consistency of the two methods for calculation of the electromagnetic mass, from the internal forces between charge elements and from the flux of field momentum, we need to show that the result in Eq. (\ref{K_s2}) also holds for $\varepsilon \rightarrow R$ and hence agrees with Eq. (\ref{m_e}) in this limit. 
 This is a little more complicated because we must now distinguish between the unit vector ${\bf n}$ in the direction from a 
charge element $ dq$  to a point on the     surface and the surface normal  ${\bf n}_{s}$ (see Fig.3). 
The momentum flux may be written as a double integral over charges $dq_1$  and $ dq_2$ at distances $\varepsilon_1$ and  $\varepsilon_2$  from a point on the sphere and with unit  vectors  ${\bf n}_1$ and  ${\bf n}_2$  towards this point. 
For simplicity we include here only the terms in Eq.(\ref{Eret_eps}) proportional to  $\varepsilon^{-2}$ and $\varepsilon^{-1}$  which determine the reactive self-force and hence the electromagnetic electron mass. Using the symmetry between $dq_1$ and $dq_2$  we obtain
$$ {\bf k}_s  \approx \frac{1}{4\pi} \int \!\! \! \int dq_1 dq_2 \left\{ \frac{ ( {\bf n}_2   {\bf n}_s) {\bf n}_1 } {\varepsilon_1^2 \varepsilon_2^2} - \frac{ {\bf n}_1 }  {2c \varepsilon_1^2 \varepsilon_2}
[(  {\bf n}_2  \dot{\btt})({\bf n}_2  { \bf n}_s) +  \right.
$$
$$
+  ({\bf n}_s   \dot{\btt})] 
- \frac{  ({\bf n}_2   {\bf n}_s) }  {2c \varepsilon_2^2 \varepsilon_1} [( {\bf n}_1   \dot{\btt})  {\bf n}_1 + \dot{\btt}] -
 \frac{  (  {\bf n}_1  { \bf n}_2 ) {\bf n}_s  }  {2 \varepsilon_1^2 \varepsilon_2^2} + 
$$
$$
\left.
 + \frac{{\bf n}_s }  {2c \varepsilon_1^2 \varepsilon_2}
[({\bf n}_2  \dot{\btt})( {\bf n}_1  {\bf n}_2) \!+\!  ({\bf n}_1  \dot{\btt})] \right\} \! \!.
\eqno(\mbox{B}1)
$$
\begin{figure} 
\begin{minipage}{\columnwidth}
\centering
\includegraphics[scale=0.6]{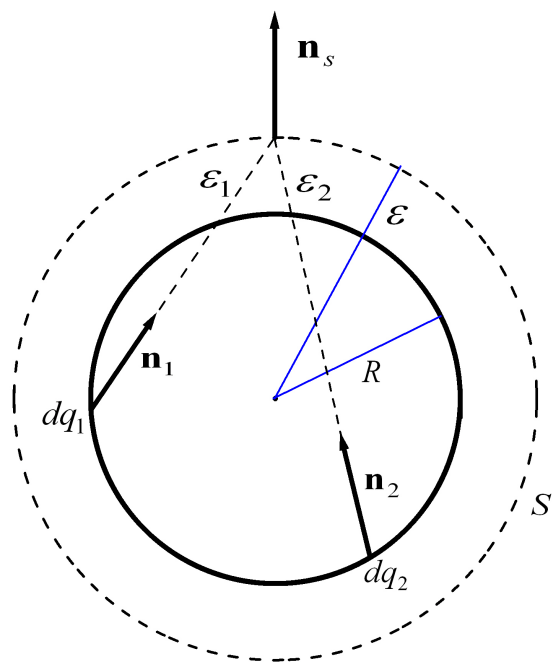}
\end{minipage}
\caption{ Geometry for calculation of the momentum flux into the sphere $S$ with radius $\varepsilon$.
The electric field is generated by a uniformly distributed charge $q$ on a spherical surface with radius $R<\varepsilon$.
}
\label{penG}
\end{figure}
This is the momentum flux into the sphere which should be integrated over the surface as in Eq.(\ref{K_s3}). For the
two terms proportional to  $(\varepsilon_1^2\varepsilon_2^2)^{-1}$   this gives zero. Consider then the second and last terms, both with the pre-factor  $(2c\varepsilon_1^2\varepsilon_2)^{-1}$  and with the factors
$$
- {\bf n}_1  [({\bf n}_2  \dot{\btt})({\bf n}_2  {\bf n}_s ) + ({\bf n}_s   \dot{\btt})] +
{\bf n}_s  [({\bf n}_2  \dot{\btt})( {\bf n}_1  {\bf n}_2) + ({\bf n}_1  \dot{\btt})].
$$
We rearrange to
$$
({\bf n}_2   \dot{\btt}) [({\bf n}_1  { \bf n}_2) {\bf n}_s   -  ({\bf n}_2   {\bf n}_s) {\bf n}_1] +
 [ {\bf n}_s ( {\bf n}_1   \dot{\btt}) -  {\bf n}_1  ({\bf n}_s   \dot{\btt})].
$$
In the expression (B1) 
the distances $\varepsilon_1$ and $\varepsilon_2$  are fixed for fixed values of $({\bf n}_1 { \bf n}_s)$  and $({\bf n}_2  {\bf n}_s)$. We keep the position of $dq_2$  fixed but average over the position of $dq_1$ on a circle around ${\bf n}_s$,  i.e., for fixed $ ({\bf n}_1  {\bf n}_s )$. This leads to
$ {\bf n}_1  \rightarrow ({\bf n}_1  {\bf n}_s ) {\bf n}_s $.  Introducing this replacement into the two expressions above we see that they both become equal to zero. This leaves
$$
{\bf k}_s  \approx \frac{1}{4\pi} \int \!\! \! \int dq_1 dq_2
\frac{  - ( {\bf n}_2   {\bf n}_s) }  {2c \varepsilon_2^2 \varepsilon_1} [( {\bf n}_1  \dot{\btt})  {\bf n}_1 + \dot{\btt}].
$$
First integrate over $ dq_2$. Only the Coulomb part of the field has survived and from electrostatics we know that the Coulomb field   from a uniformly charged spherical shell is the same outside the shell as from the total charge at the center of the sphere.  (The radius $\varepsilon$  must remain infinitesimally larger than $R$,  $\varepsilon \rightarrow R_{+}$. Within the charged surface the field is only half as large).   So the integral becomes
$$
{\bf k}_s  \approx \frac{-q}{4\pi \varepsilon^2} \int dq_1 
\frac{  1 }  {2c \varepsilon_1} [( {\bf n}_1  \dot{\btt})  {\bf n}_1 + \dot{\btt}].
\eqno(\mbox{B}2)
$$
This momentum flux should then be integrated over the sphere with radius $\varepsilon$,
\begin{equation}
{\bf K}_{s} (\varepsilon) 
= \varepsilon^2 \!\!  \int \! \!\! d\Omega_s {\bf k}_s \approx \! 
\nonumber
\end{equation}
\vspace{-5mm}
\begin{equation}
- \frac{q^2}{8\pi c} \int \!\!  \frac{d\Omega_s}{4\pi}
\int_0^{2\pi} \! \! d\varphi \! \! \int_0^\pi \! \!  d\theta \sin \theta \frac{1}{\varepsilon_1} 
\! \left[({\bf n}_1  \dot{\btt}) {\bf n}_1 \! + \dot{\btt} \right], 
\nonumber
\end{equation}
where the angles $\theta$, $\varphi$  define the direction towards $dq_1$ from the center of the sphere relative to the direction of ${\bf n}_s$.   We may perform the integration over solid angles first. For fixed values of $\theta$ and  $\varphi$  the distance $\varepsilon_1$  is constant.   The directions  ${\bf n}_s$ and ${\bf n}_1$    rotate together, covering the $4\pi$ solid  angle, so the average over ${\bf n}_s$  corresponds to an average over ${\bf n}_1$. According to Eq.(\ref{int}) we therefore obtain
$$
{\bf K}_{s} (\varepsilon)  \approx  \frac{-q^2}{8\pi c} \!\!  \int_0^{2\pi} \!\!  d \varphi  \int_0^\pi  \!\!  
\frac{d \theta \sin \theta}{  \sqrt{ (\varepsilon^2 +R^2 - 2 \varepsilon R \cos \theta ) } }   \, 
\frac{4}{3}  \dot{\btt} = 
$$
$$
= - \frac{4}{3} \frac{q^2}{2c \varepsilon}  \dot{\btt}. 
\eqno(\mbox{B}3)
$$
For $\varepsilon = R $  this result is identical to the reactive self-force in Eq.(\ref{K_s33}).

In analogy to Eq. (\ref{integration_4_3}), we must introduce the KLU-correction in the integral over forces, now with the expression (B1) for the momentum flux. 
Again the relativistic correction factor is only important for the two Coulomb terms. We
can apply the same trick as before and make the replacements   ${\bf n}_1  \rightarrow ({\bf n}_1  {\bf n}_s) {\bf n}_s$ and then the two terms can be combined. Since the scalar multiplication of ${\bf n}_1$ (or ${\bf n}_2$) by  ${\bf n}_s$ 
can be carried out after the integration over $dq_1$ (or $dq_2$) we can use the fact that the field from the charged shell is the same as that from the total charge placed at the center, and we once again obtain the correction in Eq. (\ref{integration_4_3}). 

\section*{Appendix C.  Fermi solution of 4/3-paradox}

Around 1920 Enrico Fermi wrote several papers related to the problems encountered in calculations of the
electromagnetic mass of an electron \cite{Fermi1922}. However, they have remained relatively unknown to most of the physics community, probably because the papers were published in an Italian journal. The concluding paper was also published in German and it has recently become accessible on the Internet, translated into English  \cite{Fermi1922}. In the words of Jantzen and Ruffini  \cite{Jantzen2012}, though often quoted, it has rarely been appreciated nor understood for its actual content. These authors give a detailed account of Fermi's work but like  \cite{Kalckar}  their paper is published in a journal with a limited readership, and both papers have received very few citations. We shall  here give a brief account of Fermi's approach to the problem. As seen below, there are both similarities and interesting differences to the treatments we have discussed.

There is no doubt that Fermi's view of the source of the problem was very similar to that expressed in the later paper by
Kalckar, Lindhard and Ulfbeck  \cite{Kalckar}, as demonstrated by the following quotes from the introduction. After introducing the two conflicting values or the electromagnetic mass, with and without the factor 4/3, Fermi writes:  ``Especially we will
prove: The difference between the two values stems from the fact, that in ordinary electrodynamic theory of
electromagnetic mass (though not explicitly) a relativistically forbidden concept of rigid bodies is applied. Contrary to that,
the relativistically most natural and most appropriate concept of rigid bodies leads to the value $U/c^2$   for the
electromagnetic mass.''   And further below:  ``In this paper, HAMILTON's principle will serve as a basis, being most useful for
the treatment of a problem subjected to very complicated conditions of a different nature than those considered in
ordinary mechanics, because our system must contract in the direction of motion according to relativity theory. However,
we notice that although this contraction is of order of magnitude $v^2/c^2$,  it changes the most important terms of
electromagnetic mass, i.e, the rest mass.''

Fermi's paper is not easy to read and understand, partly because he uses a description of relativistic kinematics
with an imaginary time axis (and there are a number of confusing misprints). The Lorentz transformation between reference frames in relative motion can then formally be
described as a simple rotation of the 4-dimensional coordinate system, with a complex angle of rotation. However, we
shall keep the notation applied in the main part of this paper.

The electromagnetic self-force can be derived from the principle of least action \cite{Landau}. Fermi distinguishes between two cases, A and B. In case A we disregard the relativistic effects  and consider the time  $t$  to be a common parameter for all elements of the rigid body. The part of the action responsible for the interaction of the charges with the electromagnetic field is then
$$
S_{int} = - \frac{1}{c^2}  \int A_\nu j^\nu d^4 x = - \frac{1}{c} \int dq \int_{t_1}^{t_2} A_\nu \frac{dx^\nu}{dt}dt,  
\eqno(\mbox{C}1)
$$
where  $A^\nu =(\varphi , {\bf A})$    is the  four-potential and $j^\nu =(c\rho , {\bf j})$ the  four-current, with 
$\nu = 0,1,2,3$.  The differential is  $ d^4 x = cdtdV$, where  $dV$   is a differential spatial volume. In the last expression  $x^\nu (t)$  is the world line of the charge element $dq$,   $x^\nu =(ct, {\bf r})$, and   $A_\nu$  is the four-potential at this line. 

According to the variational principle, the action should remain stationary for variations of the motion. In this connection, the definition of the integration region in Eq.(C1) is important. For a point particle, the initial and final  coordinates are to be kept fixed and this constrains the variations of the world line. Similarly, we must require that the charge elements $dq$   in Eq.(C1) have fixed  coordinates at the limits of
integration over $t$   in the variation of the action integral,
$$
\delta S_{int} = -  \frac{1}{c} \int dq \int_{t_1}^{t_2}   \left(   \frac{ \partial A_\nu}{\partial x^\mu} \delta x^\mu
       \frac{dx^\nu}{dt} +  A_\nu    \frac{d}{dt} \delta x^\nu  \right)  dt ,   
\eqno(\mbox{C}2)
$$
where $\delta x^0 =0$ while $\delta x^k$ for $k=1-3$ are arbitrary functions of $t$ except for the condition that they vanish at the limits  of integration. The last term in Eq.(C2) can be integrated by parts,
$$
\int_{t_1}^{t_2} dt  A_\nu    \frac{d}{dt} \delta x^\nu = - \int_{t_1}^{t_2} dt \frac{d A_\nu}{dt}\delta x^\nu =
$$
$$
- \int_{t_1}^{t_2} dt \frac{ \partial  A_\nu}{\partial  x^\mu} \frac{d x^\mu}{dt} \delta x^\nu .
$$
Switching the symbols $\mu$ and $\nu$  in this last term we then obtain
$$
\delta S_{int} = -  \frac{1}{c} \int dq \int_{t_1}^{t_2}   dt \left(   \frac{ \partial A_\nu}{\partial x^\mu} \delta x^\mu
  \frac{dx^\nu}{dt}   -    \frac{ \partial A_\mu}{\partial x^\nu}   \frac{dx^\nu}{dt}  \delta x^\mu  \right)  =  
$$
$$
= \frac{1}{c} \int dq \int_{t_1}^{t_2}  dt  F_{\nu \mu} \frac{dx^\nu}{dt} \delta x^\mu  = 0, 
\eqno(\mbox{C}3)
$$
where $F_{\nu \mu}$  is the electromagnetic field tensor defined in Eq. (\ref{F_mn}). When the mechanical action for the particle is included in the variation (C2), the expression (C3) provides the Lorentz force in the equation of motion for the particle.  Here we assume the mechanical mass to be zero and consider instead the balance between the force from an external  field $ {\bf E}^{(e)}$  and that from the internal field   $ {\bf E}^{(i)}$   given by Eq.(\ref{Eret_eps}) (first two terms). In the rest frame all the velocities vanish and  only the term with $\nu = 0$   remains in Eq.(C3). The four-vector $F_{0\mu}$  is given by $(0,\bf{E})$ and the variational principle leads to the relation
$$
\int {\bf E} dq  = \int \left(   {\bf E}^{(e)} + {\bf E}^{(i)} \right) dq =0 . 
\eqno(\mbox{C}4)
$$
 Fermi notes that ``we would have arrived at these equations without further ado, when we (as it ordinarily happens in the derivation of the electromagnetic mass and as it was essentially done by M. Born as well) would have assumed from the outset,  that the total force on the system is equal to zero. However, we have derived Eq. (C4) from HAMILTON's principle, to demonstrate the source of the error'' \cite{Fermi1922}.

As we have seen in Ch.2, evaluation of the internal contribution leads to the 4/3 coefficient in the self-force, 
$$
\int  {\bf E}^{(i)} dq = - \frac{4}{3} \frac{U}{c^2} {\bf g}. 
\eqno(\mbox{C}5)
$$
Here $U$ is the electromagnetic self-energy,  
$$
U= \int \!\!\! \int  \frac{dqdq'}{2\varepsilon} ,
\eqno(\mbox{C}6)
$$
where $\varepsilon$ is the distance between the two charge elements. The external force is obtained as
$$
{\bf K} = \int   {\bf E}^{(e)} dq ,
\eqno(\mbox{C}7)
$$
and we obtain from Eqs.(C4) and (C5)  the  force
$$
{\bf K} =   \frac{4}{3} \frac{U}{c^2} {\bf g} . 
\eqno(\mbox{C}8)
$$
Fermi concludes  that comparison of this equation with the basic law of point dynamics, ${\bf K}=m {\bf g}$, eventually gives us 
$$
m= \frac{4}{3} \frac{U}{c^2}. 
\eqno(\mbox{C}9)
$$ 


Fermi  then  argues that this procedure cannot be correct. 
Instead we must introduce the coordinates and the field in the momentary rest frames, i.e., in 3-dimensional planes perpendicular to the four-velocity, and specify the integration region as a section between two such planes of the world tube traveled by the body.
 In Eq.(C1), both the product of the two four-vectors and the differential  are invariant under Lorentz transformation.
We may imagine the integration region split into differential slices between two such
planes. The width of the slices is given by the differential 
time $dt$. From geometrical considerations of rigid acceleration of a body momentarily at rest, Fermi derived a relation corresponding to Eq.(\ref{dt_2_dt_1_a}) and (D9) for $\gamma =1$, 
$$
dt =  \left(1 + {\bf g \cdot R}/{c^2} \right) d t_0  ,
\eqno(\mbox{C}10)
$$
where $dt$  and $dt_0$  are the incremental  times  at  $\bf{r}$  and at a reference point on the body, ${\bf r}_0$, respectively, $\bf{g}$  is the  acceleration at ${\bf r}_0$, and $ {\bf R} = {\bf r} -  {\bf r}_0$.
 The expression  for the action integral then becomes
$$
S_{int} = - \int dq  \int \varphi    \left(1 + {\bf g \cdot R}/{c^2} \right) d \tau_0 ,
\eqno(\mbox{C}11)
$$
where $\tau_0$  is the proper time for the reference point ${\bf r}_0$ on the body.

 This defines his case B.  In his own words:  ``Now it can be immediately seen, that variation A is in contradiction with relativity theory, because it has no invariant characteristics against the world transformation, and is based on the arbitrary space $x,y,z$. On the other hand, variation B has the desired invariant characteristics, and is always based on the proper space, i.e., the space perpendicular to the world tube. Thus it is without doubt to be preferred before the previous one.'' 

 Instead of Eq. (C3) we then obtain
$$
\delta S_{int} = - \int dq  \int  \delta {\bf r} \cdot  \nabla \varphi  \,   \left(1 + {\bf g \cdot R}/{c^2} \right) d \tau_0  =
$$
$$
 =   \int dq  \int  \delta {\bf r} \cdot {\bf E}  \,   \left(1 + {\bf g \cdot R}/{c^2} \right) d \tau_0  = 0.
\eqno(\mbox{C}12)
$$

Since the displacement  $\delta {\bf r}$   is arbitrary, this leads to the relation in Eq.(C4) with the additional factor in parenthesis. This  factor  can be neglected for the external field if we choose the reference point ${\bf r}_0$ as the center of  charge (it is a very small correction in any case) and we obtain
$$
{\bf K}  =  \frac{4}{3} \frac{U}{c^2} \, {\bf g}  -  \frac{1}{c^2}
\int {\bf E}^{(i)}  ({\bf g R})    dq  . 
\eqno(\mbox{C}13)
$$
Let us consider the last term in Eq.(C13). It is only important for the leading Coulomb term in ${\bf E}^{(i)}$ and we obtain 
$$
- \frac{1}{c^2}  \int \!\!\! \int  \frac{{\bf r} - {\bf r}'}{|{\bf r} - {\bf r}'|^3}  \,  \, {\bf g} \! \cdot \! ({\bf r} - {\bf r}_0)  dqdq'  .
\eqno(\mbox{C}14)
$$
Switching notation, $ {\bf r}  \longleftrightarrow {\bf r}'$,  we obtain the same expression with the last parenthesis replaced by $({\bf r}_0 - \bf{r}')$. Averaging the two expressions we eliminate the reference coordinates  and obtain for a spherically symmetric charge distribution
$$
 -  \frac{1}{2 c^2}  \int \!\!\! \int  \frac{{\bf r} - {\bf r}'}{|{\bf r} - {\bf r}'|^3}  \, \, {\bf g} \! \cdot \! ({\bf r} - {\bf r}')  dqdq' 
=  - \frac{1}{3} \frac{U}{c^2} \, {\bf g} .
\eqno(\mbox{C}15)
$$
With this expression for the last term in Eq. (C13) we obtain the proper  relativistic equivalence
between electromagnetic mass and energy.  

Did Fermi's 1922-paper then present a satisfactory solution of the 4/3-problem, which was overlooked, not appreciated, or forgotten? In our view it fell well short of this. Fermi's argument for case B to be preferred did not identify the key problem with the calculation in case A. This calculation is not as claimed in contradiction with relativity theory and there is nothing wrong with the result in Eq.(C4), except that it is not directly relevant to the description of an accelerated rigid body. 
It expresses the condition for conserved total momentum as a function of laboratory time. 
However, as pointed out in \cite{Kalckar}, the electromagnetic mass of the rigid body is not determined by the sum of forces in Eq.(C7) through the analogy to point dynamics, leading to  Eq.(C9), but by the forces on the individual parts of the body through the relation in Eq.(\ref{Mg_0}). As expressed in Eq.(\ref{dPdt0}), this corresponds to a definition of the total momentum of the body as the sum of the momenta of its parts for fixed time in the momentary rest frame.

This relation is obtained more directly in case B because here the formulation of the variational calculation is consistent with the relativistic concept of rigid body motion.
The origin of Fermi's basic formula (C10) for time differentials at different positions is  the Eq.(4) in \cite{Fermi1922a},  which links the proper time intervals in the small spatial region in the vicinity of the world line in Riemannian space. 
In that paper Fermi also introduced the so-called   `Fermi coordinates'  applied here in case B  (see  \S 10, Ch.2  in \cite{Synge1964}).   

Fermi's paper was not cited in \cite{Kalckar}  but the authors'  view on this and later related papers is indicated by  comments to a list  of references in a note for a lecture series on `Surprises in Theoretical Physics', given by Jens Lindhard at Aarhus University in 1988 \cite{Lindhard1988}: ``E. Fermi  [1922]  started out from general relativity and suggested a covariant definition of self-mass, whereby 4/3$\rightarrow$1.  His suggestion was forgotten, even by himself. Similar attempts were tried by W. Wilson \cite{Wilson} and B. Kwal \cite{Kwal}.  Rohrlich took it up again in his textbook \cite{Rohrlich1990}, using a formal definition of a covariant classical electron. Dirac \cite{Dirac} formulated a classical theory of the electron, where he sidestepped the problem.''


\section*{Appendix D. Solution for point electron with Born-Dirac tube}

Following  Dirac \cite{Dirac},  let us surround the singular point-electron world line in space-time with  a thin tube with constant radius   $\varepsilon$  in the electron's rest frame for any instant of the electron  time-coordinate  $\tau$  in the laboratory frame. The self-force is then to be calculated from the transport of momentum across the surface of this tube. 
 Consider the four-vector $\eta^\mu = x^\mu - z^\mu (s)$,  where  $z^\mu (s) = (c\tau, {\bf r}_0 (\tau))$  is the  world line of the electron  and  $x^\mu$ is some point in the vicinity of this world line. The parameter $s$ is the proper time multiplied by the velocity of light $ds = cd\tau /\gamma$, where $\gamma (\tau) = (1 - \beta^2)^{-1/2}$ is the Lorentz factor and 
$c \btt = d {\bf r}_0 /d\tau$.  We assume that the points  $ x^\mu $ are  such that the vector $\eta^\mu$  is perpendicular to  the four-velocity of the electron,  $u^\mu = dz^\mu /ds $, and the surface of the Born-Dirac tube is then defined by two equations \cite{Dirac}, \cite{Born1909_v30}
$$
\eta^\mu \eta_\mu = - \varepsilon^2 \, ,
\eqno(\mbox{D}1)
$$
$$
\eta^\mu u_\mu = 0 .
\eqno(\mbox{D}2)
$$
These equations define a 2-dimensional structure (a sphere) in 4-dimensional space for fixed value of $s$. When
the electron moves, this structure forms a 3-dimensional surface $ f$ of a tube.  Eq.(D2)   defines a 3-dimensional
plane which is perpendicular to the four-velocity and intersects the four-sphere defined by Eq.(D1).
An analogue of the tube in three dimensions is illustrated in  Fig.4. 

Following again Dirac, let us make a variation of the point $x^\mu$  on the surface $f$  to the point $x^\mu + dx^\mu$, also on this surface. Let us suppose that this point is on the 3-dimensional plane corresponding to $ s+ds$. Differentiating the
equations (D1) and  (D2)  we obtain
$$
(x^\mu - z^\mu)(dx_\mu - u_\mu ds )=0,
\eqno(\mbox{D}3)
$$
$$
(dx^\mu - u^\mu ds ) u_\mu +(x^\mu - z^\mu) \frac {d}{ds} u_\mu ds   =0.
\eqno(\mbox{D}4)
$$
Using Eq.(D2) and the relation  $u^\mu u_\mu = 1$   we obtain from these equations the following relations
$$
\eta^{\mu} dx_{\mu} =  0,   
\eqno(\mbox{D}5)
$$
$$
u^{\mu} dx_{\mu}  = \left( 1-\eta^{\mu} \frac {d}{ds} u_\mu  \right) ds.  
\eqno(\mbox{D}6)
$$
Let us split up the four-space variation on the tube surface $f$  into a part,  $dx^\mu_\perp$,  orthogonal to the four-velocity  $u^\mu$ and a part, $dx^\mu_\parallel$,  parallel to $u^\mu$. The latter can be written as $dx_\parallel^\mu=cdt(1,\btt (\tau))$, i.e., the velocity is the same as that of the electron but the laboratory times $t$ and $\tau$ are different, as we also found from the analysis in Ch. 3.
These differentials can be visualized in the 3-dimensional analogue in  Fig.4. Here the surface $f$ is two-dimensional and  $dx^\mu$  can be split into a component along the circle, which is an intersection of the tube surface with the $x'y'$-plane, and a component parallel to the $t'$ -axis, i.e., parallel to the electron three-velocity in the laboratory  frame at the time $\tau$.

\begin{figure}
\includegraphics[scale=0.65]{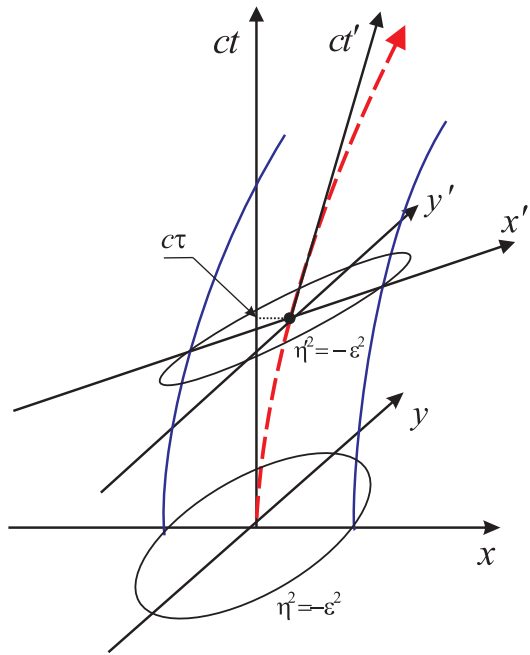}
\centering
\caption{ 
Minkowski diagram of a 3-dimensional analogue of the Born-Dirac tube around the world line of an electron (dashed red line) accelerated in the $x$-direction. Here $\tau$  is the time coordinate of the electron in the laboratory frame $(t,x,y)$ where it is at rest for $\tau=0$, $\eta^2 \equiv \eta^i\eta_i$ and $\eta'^2 \equiv \eta'^i\eta'_i$. The coordinate system $(t',x',y')$ corresponds to the rest frame at a later time $\tau$. The two circles with radius $\varepsilon$  in the $(x,y)$ and $(x',y')$  planes indicate the cuts of the tube surface with these planes and $\eta^i$  with $i = 1-3$ are the laboratory coordinates of a radius vector in one of the two circles.
 }
\end{figure}

Let us find the connection between the two times. According to Eq. (D2) the 4-plane intersecting the world-tube is defined by 
$$
c(t-\tau) = \btt \cdot ({\bf r} - {\bf r}_0 (\tau)) . 
\eqno(\mbox{D}7)
$$
The time variations of this equation gives,
$$
c(dt - d\tau) = (  {\bf R} \dot{\btt} )d\tau + \btt \cdot (d{\bf r} - c \btt d\tau), 
\eqno(\mbox{D}8)
$$
where ${\bf R} = {\bf r} - {\bf r}_0 (\tau)$.  
We found above that if we choose  $dx^\mu$  to be parallel to the electron velocity then $d{\bf r} = c\btt dt$. Insertion of  this
into Eq.(D8) leads to
$$
dt = d\tau \left(   1+ \frac{1}{c}  \gamma^2 ( {\bf R} \dot{\btt} ) \right),
\eqno(\mbox{D}9)
$$
which agrees with Eq.(\ref{dt_2_dt_1_a}).

The 3-dimensional surface element of the tube is equal to $d^3 f = |dx^\mu_\parallel| dS$,  where $dS$  is a surface element
of the sphere defined in Eqs.(D1) and (D2), 
and using the relation (D6) we find (see also the expression (66) in \cite{Dirac})
$$
d^3 f = \left(1 - \eta^\mu \frac{du_\mu}{ds} \right) ds dS  \, .
\eqno(\mbox{D}10)
$$
For a calculation of the momentum transport in the rest frame this reduces to 
$$
d^3 f  =  \left( 1 + \frac{\varepsilon }{c} \,  \, {\bf n} \cdot \dot{\btt}  \right)
c d\tau \varepsilon^2 {\bf n} d\Omega .
\eqno(\mbox{D}11)
$$
We see that the factor in the parenthesis originates in the dependence of the time differential on the spatial
coordinate in Eq.(D9), associated with the spatial variation of the
acceleration. This in turn originates in the Lorentz contraction of the rigid sphere upon acceleration.

We now obtain for the momentum transport across a section of the tube corresponding to $d\tau$, i.e., the
transport through the rigid sphere surrounding the electron corresponding to this time interval,
$$
d{\bf P}= d\tau \int \!\! \int {\bf k}_s \left(1 + \frac{\varepsilon}{c} \, \, {\bf n}_s  \cdot \dot{\btt} \right)
\varepsilon^2 d\Omega \,\, ,
\eqno(\mbox{D}12)
$$
with ${\bf k}_s$  given in  Eq. (\ref{k}). 
As we have seen   in Ch.4  this leads to complete equivalence between the electromagnetic energy and mass outside the sphere.

Dirac calculated the energy-momentum transport through the tube for the retarded field from an accelerated point charge, 
including both terms in Eq.(\ref{Eret1}). However, to obtain an equation of motion he replaced the divergent inertial self-force 
(first term in Eq.(\ref{K_s2}) but without the factor 4/3!) by a term,  
$- mc \dot{\btt}$, corresponding to a finite mass $m$ .
 He applied an expansion similar to the one discussed in Ch.2   but
more general, avoiding the assumption $\beta = 0$, and obtained a generalization of the formula (\ref{E_charge}) for the
damping force,
$$
F^\mu = \frac{2e^2}{3c} \left( \frac{d^2}{ds^2} u^\mu  + \left(  \frac{d}{ds} u^\nu  \right)^2 u^\mu  \right) ,
\eqno(\mbox{D}13)
$$
with the four-force  defined as the derivative of the four-momentum with respect to $s$.
Dirac discussed the $0'$th component of the four-force, the power term,
$$
F^0 = \frac{2e^2}{3c} \left( \frac{d^2}{ds^2} u^0  + \left(  \frac{d}{ds} u^\nu  \right)^2 u^0  \right).
\eqno(\mbox{D}14)
$$
The second term corresponds to the power of irreversible emission of radiation and, according to Dirac, gives
the effect of radiation damping on the motion of the electron. The first term is a perfect differential of a so-called
acceleration energy \cite{Schott} and corresponds to reversible exchange of energy with the near field (see
also \cite{Schwinger}). However, it should be noted that the other terms of the four-force  do not separate so neatly and are mixed under Lorentz transformations.

An interesting derivation of the formula (D13) is given in  \cite{Landau}  (see also \cite{Pauli} \S 32). The first term is an obvious relativistic generalization of Eq.(\ref{E_charge}) but it does not have the property required by any four-force  that it be perpendicular to the four-velocity.  The second term is then added as a plausible extension remedying this deficiency. And  it is this term that now accounts for the radiation reaction!

\end{document}